\documentclass[10pt,letterpaper]{article}
\usepackage[top=0.85in,left=2.75in,footskip=0.75in]{geometry}

\usepackage{amsmath,amssymb, amsthm}

\usepackage{changepage}

\usepackage[utf8]{inputenc}

\usepackage{textcomp,marvosym}

\usepackage{cite}

\usepackage{nameref,hyperref}

\usepackage[right]{lineno}

\usepackage{multirow}
\usepackage{microtype}
\DisableLigatures[f]{encoding = *, family = * }

\usepackage[table]{xcolor}

\usepackage{array}

\usepackage{subcaption}
\newcolumntype{+}{!{\vrule width 2pt}}

\newlength\savedwidth

\raggedright
\usepackage[aboveskip=1pt,labelfont=bf,labelsep=period,justification=raggedright,singlelinecheck=off]{caption}

\makeatletter
\renewcommand{\@biblabel}[1]{\quad#1.}
\makeatother

\usepackage{lastpage,fancyhdr,graphicx}
\usepackage{epstopdf}
\fancyheadoffset[L]{2.25in}
\fancyfootoffset[L]{2.25in}
\def\BibTeX{{\rm B\kern-.05em{\sc i\kern-.025em b}\kern-.08em
    T\kern-.1667em\lower.7ex\hbox{E}\kern-.125emX}}

\definecolor{mypink2}{RGB}{219, 48, 122}
\definecolor{mypurple}{RGB}{51,51,178}
\definecolor{myred}{RGB}{150,50,50}
\definecolor{myred2}{RGB}{255,10,10}
\definecolor{mygreen}{RGB}{50,150,50}

\usepackage[normalem]{ulem}

\begin{document}
\vspace*{0.2in}

\begin{flushleft}
{\Large
\textbf\newline{RegRole: Regularized Role Detection and Prediction in Temporal Dynamic Networks} % Please use "sentence case" for title and headings (capitalize only the first word in a title (or heading), the first word in a subtitle (or subheading), and any proper nouns).
}
\newline
\\
Emily J. Evans\textsuperscript{1\Yinyang},
Weihong Guo\textsuperscript{2\Yinyang},
Carlotta Domeniconi\textsuperscript{3,\Yingyang}
\\
\bigskip
\textbf{1} Department of Mathematics, Brigham Young University, Provo, UT, USA
\\
\textbf{2} Department of Mathematics, Case Western Reserve University, Cleveland, OH, USA
\\
\textbf{3} Department of Computer Science, George Mason University, Fairfax, VA, USA
\\
\bigskip

% Insert additional author notes using the symbols described below. Insert symbol callouts after author names as necessary.
% 
% Remove or comment out the author notes below if they aren't used.
%
% Primary Equal Contribution Note
\Yinyang These authors contributed equally to this work.

% Use the asterisk to denote corresponding authorship and provide email address in note below.
* ejevans@mathematics.byu.edu

\end{flushleft}
\section*{Abstract}
This paper introduces a dynamic role discovery technique in temporal dynamic networks, utilizing temporally regularized Non-negative Matrix Factorization (NMF). Our technique differs from existing dynamic role analysis techniques by creating a consistent set of roles across all time periods, as well as a universal transition matrix that describes the probability of transitioning between roles.  We also apply a regularization penalty to ensure that role membership does not change dramatically between time periods making our model more robust against real-world noise. We test our data on five real-world and one synthetically simulated dataset using both engineered and automatically generated features.  We demonstrate that the proposed regularized role detection method, for appropriate regularization weight parameter  reduces prediction errors compared to other techniques. Furthermore, trace analysis of the transition matrices indicates that our method yields a more stable system, that is, individuals are more likely to stay in their roles with fewer arbitrary transitions. Our model learns time-aligned roles, captures behavioral transitions over time, and scales efficiently to large and sparse graphs.

\section*{Introduction}

Many real-world systems—including communication platforms, online collaboration and software repositories, financial transaction systems, and transportation or mobility infrastructures—are naturally represented as \emph{dynamic graphs}, in which both nodes and edges evolve over time \cite{HolmeSaramaki2012}. In such settings, the network structure is not static but reflects ongoing processes such as user arrival and departure, changing interaction patterns, and evolving functional responsibilities of nodes.

Beyond the basic question of \emph{who is connected to whom}, a central objective in network analysis is to understand \emph{how} nodes participate in the system. In particular, nodes may assume characteristic \emph{structural or behavioral roles}, such as brokers mediating information flow, authorities attracting many inbound interactions, peripheral followers with limited connectivity, or coordinators linking multiple substructures. Importantly, these roles are not defined solely by local connectivity, but by higher-order interaction patterns that capture how nodes relate to the broader network context.
This perspective contrasts with \emph{community detection}, which groups nodes based on dense intra-group connectivity. 

In role discovery, nodes are instead grouped according to similar \emph{patterns of interaction}, regardless of whether they are directly connected to one another \cite{RossiAhmed2015}. For example, two nodes may both serve as bridges between otherwise distant regions of the graph, or may each maintain many outgoing links to high-degree hubs, even if they never interact directly. As a result, role discovery offers a complementary lens on network organization, emphasizing functional equivalence over topological proximity.

Identifying and tracking roles in dynamic graphs underpins a wide range of downstream tasks. These include anomaly and change-point detection (e.g., identifying nodes whose behavior deviates from their historical role), user lifecycle modeling, intervention and policy design, network summarization, and forecasting of future structural evolution \cite{Rossi2012RoleDynamics,Rodrigues2025PredictingDynamics}. In many applications, roles provide a more stable and interpretable abstraction than raw links, especially in large-scale or noisy temporal networks.

\section{Related Work}

\subsection{Role discovery as representation learning}

Role discovery is commonly formulated as an \emph{unsupervised representation learning} problem. Given a graph (or sequence of graphs), the goal is to induce a compact set of latent role types and to assign each node a representation over these roles, often allowing \emph{mixed membership} so that nodes may simultaneously express multiple functional patterns \cite{RossiAhmed2015}. Early approaches relied on matrix factorization and feature-based decompositions, while more recent methods leverage neural embeddings and deep learning frameworks. When extending role discovery to dynamic networks, two fundamental challenges arise.
First, \emph{non-stationarity}: roles themselves may evolve over time. New roles can emerge as the system grows, existing roles may dissolve, and the semantics of roles may drift due to changes in interaction patterns or external conditions. Capturing such evolution requires models that are flexible enough to adapt to genuine structural change, rather than forcing static assumptions onto dynamic data \cite{Rossi2012RoleDynamics, Pei2018DyNMF}. Second, \emph{temporal alignment}: role identities must remain comparable across time windows. Without explicit alignment, independently learned role representations may suffer from arbitrary permutations or fragmentation, making longitudinal analysis difficult or meaningless. Effective dynamic role discovery therefore requires balancing \emph{adaptability} (to reflect real change) with \emph{stability} (to preserve interpretability and comparability over time).

\subsection{Role dynamics and temporal modeling}

Several methods have been proposed to explicitly address role evolution in temporal networks. The Role-Dynamics framework \cite{Rossi2012RoleDynamics} introduced efficient techniques for mining large dynamic graphs by incrementally updating role representations over time. DyNMF \cite{Pei2018DyNMF} further extended this idea using dynamic non-negative matrix factorization, enabling smooth temporal transitions in role memberships.

More recent work has explored the relationship between \emph{structural role embeddings} and \emph{proximity-based embeddings}, highlighting that standard graph embedding methods optimized for link prediction or community structure may fail to capture role equivalence \cite{Rossi2020ProximityRoles}. This distinction has motivated the development of role-aware embedding techniques tailored to functional similarity rather than neighborhood overlap.

In parallel, advances in \emph{temporal graph neural networks} have enabled richer modeling of time-evolving network structure \cite{Zheng2025DynamicGNNSurvey}. Role-based Temporal GCNs (RTGCN) \cite{Du2024RTGCN} explicitly incorporate role information into temporal message passing, while other approaches leverage dynamic embeddings for link prediction and forecasting \cite{Wu2023EnSAT, Mei2024SciRep, Zhu2023NewNodeTemporal}. Although many of these models are not explicitly framed as role discovery methods, they underscore the growing importance of temporally aligned representations in dynamic graph analysis.

\subsection{Recent advances and broader context}

Beyond role-specific methods, related advances in unsupervised and self-supervised learning for temporal graphs have influenced the field. Contrastive learning frameworks for dynamic representations \cite{jiao2024contrastive} provide powerful tools for capturing temporal dependencies without labeled data. Work on predicting dynamical network evolution \cite{Rodrigues2025PredictingDynamics} and on jointly modeling community structure and anomalies \cite{Safdari2024CommunityAnomaly} further highlights the need for representations that are both temporally coherent and structurally meaningful. {Recent benchmarking efforts have also emphasized realistic temporal splits,
large-scale evaluation, and reproducibility across node- and edge-level prediction tasks \cite{Huang2023TGB}. Extensions of the Temporal Graph Benchmark to heterogeneous and
multi-relational networks further demonstrate that scalability and fair evaluation remain unresolved issues for temporal graph models \cite{Gastinger2024TGB2}.}

Taken together, these lines of work emphasize that dynamic role discovery is not merely an extension of static role analysis, but a distinct and challenging problem at the intersection of temporal modeling, representation learning, and network science. Effective solutions must jointly account for structural equivalence, temporal evolution, and semantic stability—requirements that remain active areas of research.

\begingroup

A notable example is given by the RolX framework~\cite{RolX12}, which combines recursive structural features with 
nonnegative matrix factorization (NMF) {and K-Means clustering} to extract possibly interpretable roles. Extensions to temporal settings regularize 
factorizations across time, ensuring smooth transitions in role memberships while preserving consistent role semantics~\cite{Rossi2012RoleDynamics, rossi2013modeling} 
. In contrast, we utilize NMF and a consistent set of roles across time.  When we use engineered structural features, this set of roles is easily interpretable, and when we utilize recursive structural features the roles are not as easily interpretable, but it is still possible to interpret them.  In our work we also {introduce a unified transition matrix} and  a regularization {so that role membership at one time multiplied by the transition matrix is close to the role membership at the next time}, resulting in a more realistic situation under the influence of noise.

%{\color{red}{We need to say something on how we are different and better than Rossi et al.}}

We introduce a dynamic topic discovery framework named RegRole, for role detection and prediction, based on temporally regularized NMF. Our method 
(i) learns time-aligned role topics over structural and behavioral features, 
(ii) induces node-level role mixtures that evolve with controlled smoothness, and 
(iii) scales to large and sparse graphs. 
Empirically, it delivers more stable role semantics across time, sharper change detection, and improved downstream prediction 
compared to {DyNMF for both engineered features and automatically generated (i.e., structurally based) features.  Moreover, our method has a predictive ability due to universal role and transition matrices that other techniques do not have.}

\section{Background}

To facilitate the discovery of roles in networks, the first step is often to convert the network into a feature matrix $X$, where each row corresponds to a node in the network.  These features matrices can be either engineered, or automatically generated (see Section~\ref{sec:feature_gen} and \cite{REFEX, Revelle16}).   Non-negative matrix factorization was first used for the task of discovering roles of nodes from these feature matrices in \cite{RolX12}. It is computationally efficient and non-negative factors simplify the interpretation of roles and memberships. In  \cite{RolX12}, only static networks are considered and features correspond to aggregated per-node structural attributes \cite{Henderson11}. The resulting node-feature matrix is then decomposed using NMF, and the emerging basis vectors give the node roles in the input network. 

Given a node-feature matrix $X \in \mathbb{R}^{m \times k}$ with $m,k$ denote the number of nodes and features respectively, the RolX algorithm in  \cite{RolX12} generates a rank-$r$ approximation $UV \approx X$, using NMF. Here each row of $U \in \mathbb{R}^{m \times r}$ represents a node’s fractional component in an $r$ dimensional subspace, and each column of $V \in \mathbb{R}^{r \times k}$ specifies how membership in each of these dimensions contributes to estimated feature values. 

Formally, NMF seeks two non-negative low-rank matrices $U$ and $V$ to solve the following optimization problem:
\begin{equation}
\underset{U,V}{\arg\min} \ \|X - UV\|_F^2 \quad \text{subject to } U \geq 0, \ V \geq 0,
\label{eq:NMF}
\end{equation}
where $\|\cdot\|_F$ denotes the Frobenius norm. The nodes are then clustered based on the rows in $U$ using K-Means to determine the final roles. 
%There are many methods to generate such an approximation (e.g., SVD, spectral decomposition), and RolX\textcolor{red}{Do we want to refer to ROLX here?} is not tied to any particular approach. 

%For this \WG{Emily: does "this" mean RolX or our proposed? It is not clear to me.} study, we also use Non-negative Matrix Factorization to decompose our feature matrix because it is computationally efficient and non-negative factors simplify the interpretation of roles and memberships. 

The authors in \cite{DBMM13} apply NMF for modeling dynamic behavior in evolving graphs. They identify roles in each graph snapshot individually, and then use the resulting node-role matrices in consecutive snapshots to compute a role transition matrix. Let $X_t \in \mathbb{R}^{m\times k}$ be the raw node-role feature matrix at  time $t = 0, 1, \cdots n$, NMF algorithm 

\[
\underset{U_t,V}{\arg\min} \ \|X_t - U_tV\|_F^2 \quad \text{subject to } U_t \geq 0, \ V \geq 0,
\label{eq:NMF2}
\]
is used multiple times to obtain membership matrix $U_t$ at different time $t$ and a unified role matrix $V$. A transition matrix $T$ represents how likely a node is to transition from one role to another for that particular time interval is estimated using NMF such that $U_{t-1}T \approx U_t$. Variants of the idea include stacked transition which stacks the training examples from the $k$ previous time steps and finds $T$ so that 

\[
\begin{bmatrix}
U_{t-1} \\
U_{t-2} \\
\vdots\\
U_{t-k} 
\end{bmatrix}T \approx
\begin{bmatrix}
U_{t} \\
U_{t-1} \\
\vdots\\
U_{t-k+1} 
\end{bmatrix}.
\]

Such transition matrix is then used to analyze the dynamic behavior of the graph and to predict changes in nodes' roles. The focus of this work is on modeling structural changes of nodes over time, rather then on discovering interpretable and meaningful roles. 
%This approach computes a different role-feature matrix for each snapshot, which are then discarded.  

The authors in \cite{Revelle16} introduced a NMF-based methodology for discovering and tracing roles over time in dynamic networks. The network is partitioned into snapshots, nodes' structural and behavioral features are extracted from each
snapshot, and roles are found independently in each snapshot. The authors discover roles which are persistent across the snapshots and common among different networks.

DyNMF \cite{DyNMF18} is a NMF-based approach that models roles and role transitions simultaneously: at each time step $t$, DyNMF solves the following problem to finds a time dependent transition matrix $M_t$, in addition to $U_t, V_t$:

\begin{equation}
\begin{aligned}
\underset{U_t,V_t,M_t}{\arg\min} \quad 
& \|X_t - U_tV_t\|_F^2 + \|X_t - U_{t-1}M_tV_t\|_F^2 \\
\text{subject to} \quad 
& U_t \geq 0,\quad V_t \geq 0,\quad M_t \geq 0.
\end{aligned}
\label{eq:DyNMF}
\end{equation}

\noindent $M_t$ captures the transitions between previous roles and current roles. One can extend the above first-order assumption to any higher-order version \cite{akoglu2015graph,kent-2015-cyberdata1}.

\section{Methodology of the Proposed Model}
%\subsection{Regularized Role Detection}

\subsection{RegRole: Regularized Role detection and prediction}
Let $X_t \in \mathbb{R}^{m\times k}$ be the raw feature matrix at different time $t = 0, 1, \cdots n$ with $m,k$ denote the number of nodes and features respectively.  We require the number of nodes and the number of features to remain constant throughout all of the time periods.  We aim to find $U_t \in\mathbb{R}^{m\times r}$,  $V \in \mathbb{R}^{r \times k}, M \in \mathbb{R}^{r\times r}$ with $r$ the number of roles, so that $X_t \approx U_t V, t = 0,1,\cdots n$ and $U_{t-1}M \approx U_t, t = 1, 2, \cdots n$.  $M$ is meant to be a transition matrix from one time snapshot to the next and can be used to predict role weights at the next time steps. 

The problem under consideration is as follows
\begin{equation}
\min_{U_t, V,M\geq 0} \sum_{t=1}^N\left( \frac{1}{2}\|X_t- U_tV\|_F^2 + \frac{\beta}{2} \|U_{t-1}M-U_t\|_F^2\right) +
\frac{1}{2}\|X_0-U_0V\|_F^2
\label{model}
\end{equation}
We stack all $X_t, t = 0, 1, \cdots n$ vertically in order and form a matrix $X\in \mathbb{R}^{nm\times k}$ and do the same thing to stack $U_t, t = 0, 1, \cdots n$ vertically in order and form a matrix called $U\in \mathbb{R}^{nm \times r}$. We also define

$A_1 =
\begin{pmatrix}
    I & 0 & 0 & 0& 0\\
    0 & I & 0 & 0  & 0\\
    0 & 0 & I & 0 & 0\\
    \vdots & \vdots  & \ddots &\vdots &\vdots\\
    0 & 0 & \hdots & I & 0
\end{pmatrix}, A_2 =
\begin{pmatrix}
    0 & I & 0 & 0& 0\\
    0 & 0 & I & 0  & 0\\
    0 & 0 & 0 & I & 0\\
    \vdots & \vdots  & \ddots &\vdots &\vdots\\
    0 & 0 & \hdots & 0 & I
\end{pmatrix}  $
where $I, 0$ represents $m\times m$ identity and zero matrix respectively. Model $(\ref{model})$ can then be reformulated in a neater form:

\begin{equation}
\min_{U, V,M\geq 0} \mathcal{E}(U,V,M) = \frac{1}{2}\|X- UV\|_F^2 + \frac{\beta}{2} \|A_1UM-A_2U\|_F^2
\label{model2}
\end{equation}

We first rewrite $\mathcal{E}(U,V,M)$ in trace form as

% \begin{multline*}\mathcal{E}(U,V,M,\Lambda_1,\Lambda_2,\Lambda_3) = -\tr(X^TUV) + \frac{1}{2}\tr(V^TU^TUV) + \beta\tr(M^TU^TA^T_1A_1UM) \\-2\beta\tr(M^TU^TA_1^TA_2U +\beta\tr(U^TA_2^TA_2U) \end{multline*}

% It is straightforward to see
% \begin{multline*}\frac{\partial\mathcal{E}}{\partial U} = -XV^T + UVV^T + 2\beta A_1^TA_1UMM^T \\- 2\beta(A_2^TA_1UM + A_1^TA_2UM^T)
% +2\beta(A_2^TA_2U) 
% \end{multline*}

% \[\frac{\partial \mathcal{L}}{\partial V} = -U^TX +U^TUV \]
% and
% \[\frac{\partial\mathcal{L}}{\partial M} = 2\beta U^TA_1^T A_1UM - 2\beta U^TA_1^TA_2 U .\]

\begin{align*} \mathcal{E}(U,V,M,\Lambda_1,\Lambda_2,\Lambda_3) ={}& -\operatorname{tr}\!\left(X^{T}UV\right) + \frac{1}{2}\operatorname{tr}\!\left(V^{T}U^{T}UV\right) \\ &+ \beta \operatorname{tr}\!\left(M^{T}U^{T}A_1^{T}A_1UM\right) \\ &- 2\beta \operatorname{tr}\!\left(M^{T}U^{T}A_1^{T}A_2U\right) \\ &+ \beta \operatorname{tr}\!\left(U^{T}A_2^{T}A_2U\right). \end{align*} It is straightforward to see that \begin{align*} \frac{\partial \mathcal{E}}{\partial U} ={}& -XV^{T} + UVV^{T} + 2\beta A_1^{T}A_1UMM^{T} \\ &- 2\beta\left( A_2^{T}A_1UM + A_1^{T}A_2UM^{T} \right) \\ &+ 2\beta A_2^{T}A_2U, \end{align*} \begin{align*} \frac{\partial \mathcal{E}}{\partial V} &= -U^{T}X + U^{T}UV, \end{align*} and \begin{align*} \frac{\partial \mathcal{E}}{\partial M} &= 2\beta U^{T}A_1^{T}A_1UM - 2\beta U^{T}A_1^{T}A_2U. \end{align*}

 We adopt the method of Lee~\cite{lee2000algorithms} to obtain a multiplicative update shown below:

            \[U\leftarrow  = U\circ\frac{XV^T}{UVV^T+2\beta A_1^TA_1UMM^T+2\beta A_2^TA_2U-2\beta(A^T_2A_1UM+A_1^TA_2UM^T)}\]

            \[M \leftarrow M\circ \frac{U^TA_1^TA_2U}{U^TA_1^TA_1UM}\]
           
           \[V \leftarrow V\circ \frac{U^TX}{U^TUV}.\]
We enforce non-negativity of the iterates by projecting $U,V,M$ to be nonnegative through forcing negative entries of each to be 0. 
          Note that if an entry in the denominator is zero we set it equal to a small number such as $10^{-8}$ to avoid dividing by 0.  Note $\circ$ indicates entry-wise multiplication, and the displayed division sign indicates entry-wise division.

 \subsection{Role prediction}

 Our model has a natural role prediction capability: multiplying the membership matrix $U_t$ by the universal transition matrix $M$, one gets an estimation of the membership matrix $U_{t+1}$ at the next time step and one can keep multiplying by $M$ to predict further timesteps. Unlike DyNMF~\cite{DyNMF18} our model seeks to identify a universal set of roles, in a manner similar to the work of~\cite{Revelle16}.  We also identify a universal transition matrix that holds across all of the time periods, instead of identifying a transition matrix across each time period.  The reason for these two characteristics of our model is so that we can predict the role characteristics of an individual in the future.

 \begin{table}
 \begin{tabular}{|l|l|}
 \hline
 Feature&Short Description\\
 \hline
 
     In-degree& Count of incoming edges\\
Out-degree&Count of outgoing edges\\
Weighted in-degree&Count of incoming interactions\\
Weighted out-degree& Count of outgoing interactions\\
Reciprocity& Ratio of reciprocated edges over all \\
&outgoing edges\\
New activity count& Count of new outgoing edges\\
Social strategy& Ratio of new outgoing edges over all\\ &outgoing
edges\\ 
Betweenness centrality&Number of all shortest paths which\\ &pass through the node\\
PageRank& PageRank measure of centrality\\
Weighted PageRank&Weighted variant of PageRank\\
Transitivity&Probability any two neighbor nodes are\\ &connected\\
Weighted transitivity&Weighted variant of transitivity\\
\hline
 \end{tabular}
     \caption{A description of the engineered features used.}
    \label{tab:engfeature}
 \end{table}

\section{Experiments}

\subsection{Data}\label{sec:feature_gen}
Our data consists of five real-world data sets and one set of synthetic data, the statistics of which are displayed in Table~\ref{tab:datasetinfo}. Four of the real-world datasets (Enron, Facebook, Reality and Slashdot) are from the network data repository~\cite{nr}.  The Scratch dataset is the one used in~\cite{Revelle16}.  

The synthetic dataset was created in a similar manner to that in~\cite{DyNMF18}.  More specifically, we created the synthetic dateset from a series of graph time snapshots each of which had four distinct roles, clique nodes, star-edge nodes, star-center nodes, and bridge nodes, which bridge between different node types.  Our synthetic data was initialized by creating four cliques with an average of 11 nodes and a standard deviation of 1, 3 stars with an average of 12 nodes, again with standard deviation 1, and twenty bridge nodes.  We then generated 20 time snapshots with the probability of nodes changing roles of 2 percent.  This meant that on average two nodes change roles at every time snapshot.

In our experiments we utilized both engineered features and automatically generated features.  The engineered feature matrices contain the features described in Table~\ref{tab:engfeature}, and are consistent with the features used in~\cite{Revelle16}.  For each dataset we also generated automatic features using the ReFeX algorithm~\cite{REFEX}. To make sure the number of features is consistent across time periods, we ran ReFeX on a large network, that spanned across all of the time periods.  We did this by duplicating the nodes, i.e., the nodes appeared $n$ times where $n$ is the number of time periods.  We then created connections among identical nodes between time periods.  After running ReFeX, we were able to have feature matrices that were consistent across time periods. The number of engineered features and the number of automatically generated features are shown in Table~\ref{tab:datasetinfo}. For the scratch dataset we were only able to obtain the engineered features and hence this dataset is not included in our experiments with automatically generated features.  

\begin{table}[]
    \centering
    \begin{tabular}{|l|c|c|c|c|}
        \hline
         Dataset& Nodes & Edges & Snapshots &Automated \\
         & & & & Features\\
         \hline
         Scratch&5934 &Unknown &19 & N/A\\
         Enron&151 & 50572&82 & 48\\
         Facebook&46952 & 106940&7 & 89\\
         Reality&6809 &16479 &8 & 69\\
         Slashdot&51068 &130324 &6 & 78\\
         Synthetic & 100 &6576 & 20& 12\\
         \hline
    \end{tabular}
    \caption{Data characteristics (number of nodes, edges, time snapshots and automated features) of six different data sets (Enron, Facebook, Reality, Slashdot, Scratch and synthetic) used for numerical results comparison. }
    \label{tab:datasetinfo}
\end{table}
%\WG{Put in summary statistics (e.g., number of nodes and edges.  See Rossi for ideas. including availability in a table}The synthetic dataset consists of a collection of nodes that we define as being clique nodes, bridge nodes, or star nodes.  % We will experiment on different feature matrices (engineered feature matrix and automatically determined feature matrices (Rossi)).
%We will test the performance of role detection, role prediction when the value of the parameter $\beta$ varies. We will compare with DyNMF, our own work, Matt's work for role detection and role prediction. 

\subsection{The choice of the rank $r$ and regularization constant $\beta$}

{There are two parameters in the proposed model RegRole: one is the rank $r$ (i.e., the number of roles to discover) and the other one is the regularization constant $\beta$.
 When determining the best value of $r$ we experimented with several different values and no regularity to determine the optimal value. Setting the rank $r$ at 6 worked well for all of the datasets, so we fix $r$ to be 6.} Regarding regularity constant $\beta$,  we tested with a large variety of $\beta$ values. For each experiment and each data set we started with a set of unnormalized $\beta$ values  \{0,0.005,0.015, 0.025,0.05, 0.075,0.1,0.15, 0.2, 0.25,0.3, 0.5,0.8,1,2,3,10\}.  We then normalized the $\beta$ values by multiplying by the proposed rank $r$ (in our case 6) and dividing by the number of features to account for size difference of matrices in the two terms of the proposed model.  In the tables we display values of $\beta$ that produced reasonable results for all of the data sets, so that we can illustrate the utility of the method.  For the figures, which illustrate the results of our method on the Facebook dataset specifically, we chose to display only three $\beta$ values so as not to clutter the plots.  These values were chosen because they resulted in good experimental results (i.e., lower error values with a known ground truth).  The optimal value of $\beta$ varied with the choice of dataset, which indicates that the choice of regularization parameter should be tuned with the dataset.
\subsection{Experiment 1: The effect of $\beta$ on role detection.}
Before considering the utility of our model to predict roles, we first investigate the effect of changing the value of $\beta$, our regularization parameter, on the detection of the roles. In Fig~\ref{fig:rolecomparison},  we display the consequences of changing the value of $\beta$ for the Scratch dataset. The color of the bars correspond to the different role vectors resulting from the proposed model with different $\beta$ values. For each fixed color (fixed $\beta$ value), the entries of $V$ corresponding to each of the six ($r = 6$) roles (Explorer, Reciprocated, Friendly, Regular, Community Member, Other) detailed in~\cite{Revelle16}.  The height of each bar in each subfigure tells the value of $V_{jk}$ for the $j$-th role and the $k$-th feature. Features are listed on the  horizontal axis of the last two subfigures). Note, the proposed model with $\beta = 0$ is equivalent to model in ~\cite{Revelle16}. The roles found are similar and as expected, as the value of $\beta$ increases, regularized role detection spreads the weight given to each feature out among all of the features.

\begin{figure}
     \centering

         \centering
         \includegraphics[width=.75\textwidth]{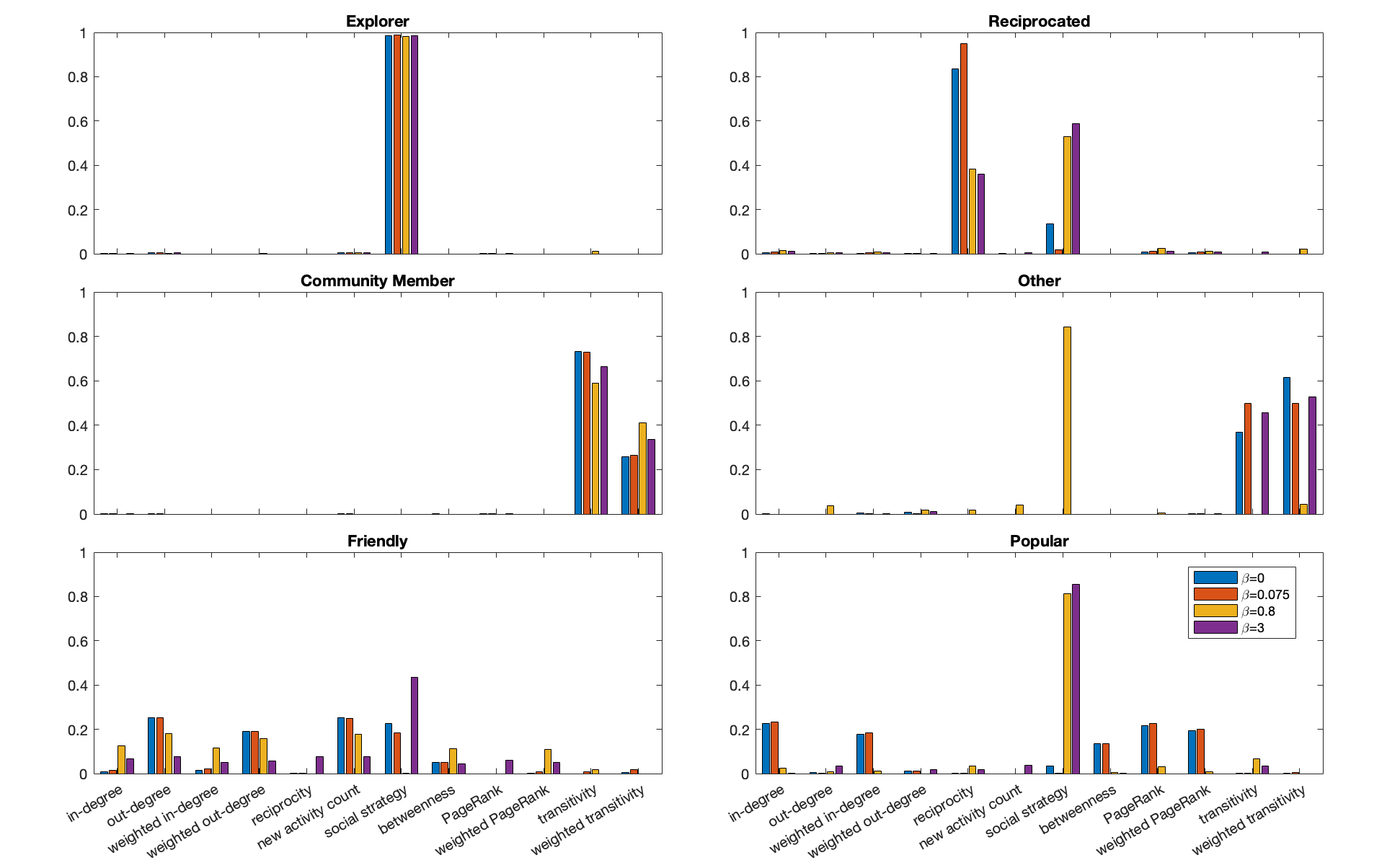}
         \caption{Comparison of the roles found from the Scratch dataset using the proposed RegRole with different $\beta$ values and the averaged roles found  from~\cite{Revelle16} (equivalent to the proposed model with $\beta = 0$.) The roles found are similar, however RegRole spreads the weight given to particular features in nearly all roles. The names of the roles are taken from~\cite{Revelle16}.}
        \label{fig:rolecomparison}
     \end{figure}

%Update code/Figure RoleComparison)
%Figure set 1: Role detection comparison for each of the data sets. Choose the best results chose from different $\beta$ values.

%Figure set 2: Prediction error: synthetic (all 5 data sets), ours, DyNMF ()

%Figure set 3: role prediction error comparison

\subsection{Experiment 2: The predictive power of the transition matrix}  This experiment mimics the work done by~\cite{DyNMF18} to analyze the predictive power of the transition matrix to determine the role distribution at the next step.  In this experiment we utilize the entirety of each data set to determine appropriate values for the $U_t, V, M$ ({the proposed} regularized role transition matrix) and $U_t, V_t, M_t$ (DyNMF).  We display the results for the Facebook dataset in Fig~\ref{fig:facebookallpredict}.  For the DyNMF curves, we follow the same procedure as in~\cite{DyNMF18}.  One of the curves calculates $||U_t- U_{t-1}M_{t-1}||$ ({denoted as \textit{previous}} in the caption of Fig. \ref{fig:facebookallpredict}) and the second curve displays the norm of the difference with the average transition matrix over the prior time steps, i.e., $||U_t-U_{t-1}\left(\frac{1}{t-1}\sum_{i=1}^{t-1} M_i\right)||$ ({denoted as \textit{average}} in the caption of Fig. \ref{fig:facebookallpredict}).  For the RegRole curves we simply display $||U_t-U_{t-1}M||$ since this method results in a universal transition matrix. As can be seen in the figure, RegRole results in a consistently and significantly lower predictive error.  In Tables~\ref{tab:rta} and ~\ref{tab:rta_AUTO} we display the max, min, and average prediction results for both RegRole and DyNMF using engineered and automatically generated features respectively for all six datasets. 
{From Table~\ref{tab:rta} and the left panel of
Fig~\ref{fig:facebookallpredict}, one can see that the proposed RegRole method is consistently better than DyNMF in role detection using engineered features: RegRole yields lower maximum, minimum, and
mean error measurements. For automatically generated features, however, the prediction errors reported in
Table~\ref{tab:rta_AUTO} are generally much larger than those reported in Table~\ref{tab:rta}. This implies that the assumptions of the proposed RegRole model fit the engineered features well, but may not fit the automatically generated features as consistently. The right panel of Fig~\ref{fig:facebookallpredict} shows that DyNMF performs better on the Facebook data with automatically generated features.}
This trend is observed for several, but not all, datasets, as shown in Table~\ref{tab:rta_AUTO}.
{However, for some values of $\beta$ and some datasets, RegRole still outperforms DyNMF. All other comparisons following also show that RegRole outperforms DyNMF.}

\begin{figure}[htbp]
  \centering
 \includegraphics[width=0.48\textwidth]{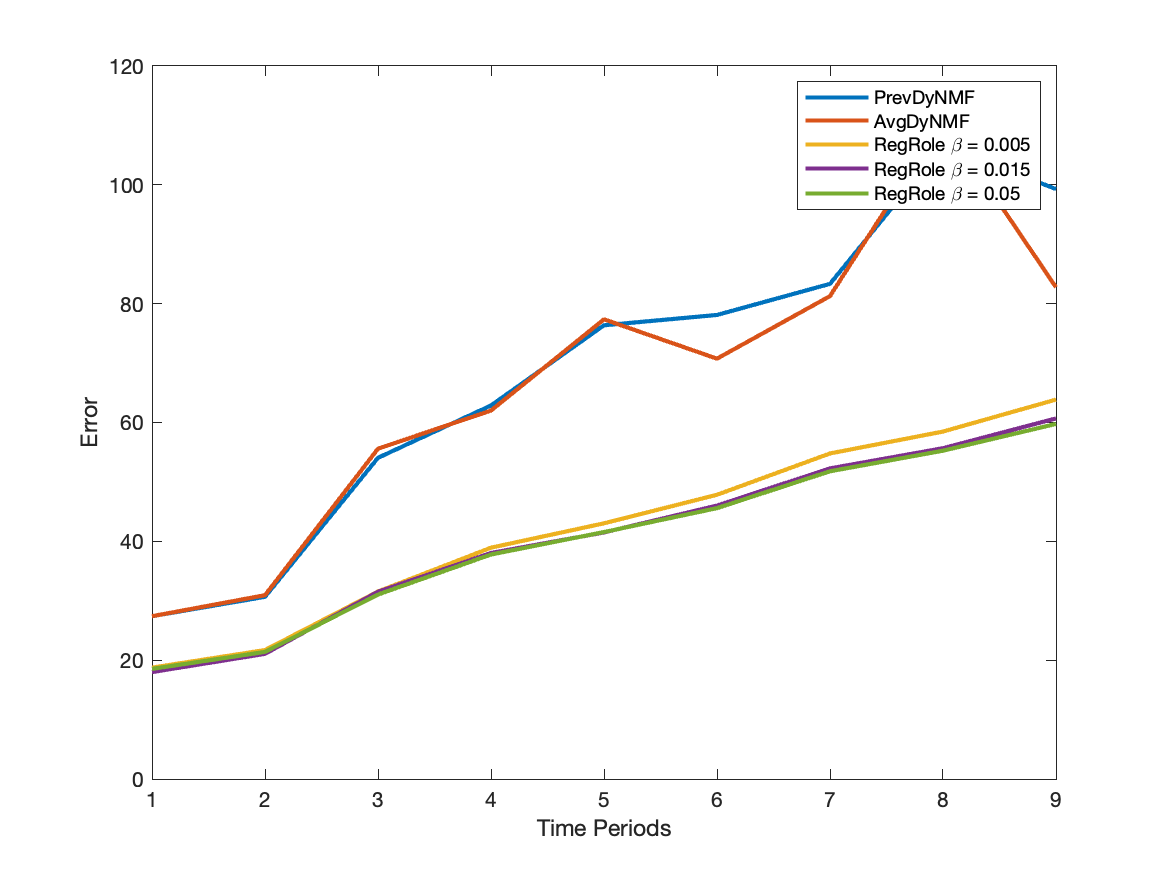}\hfill
  \includegraphics[width=.48\linewidth]{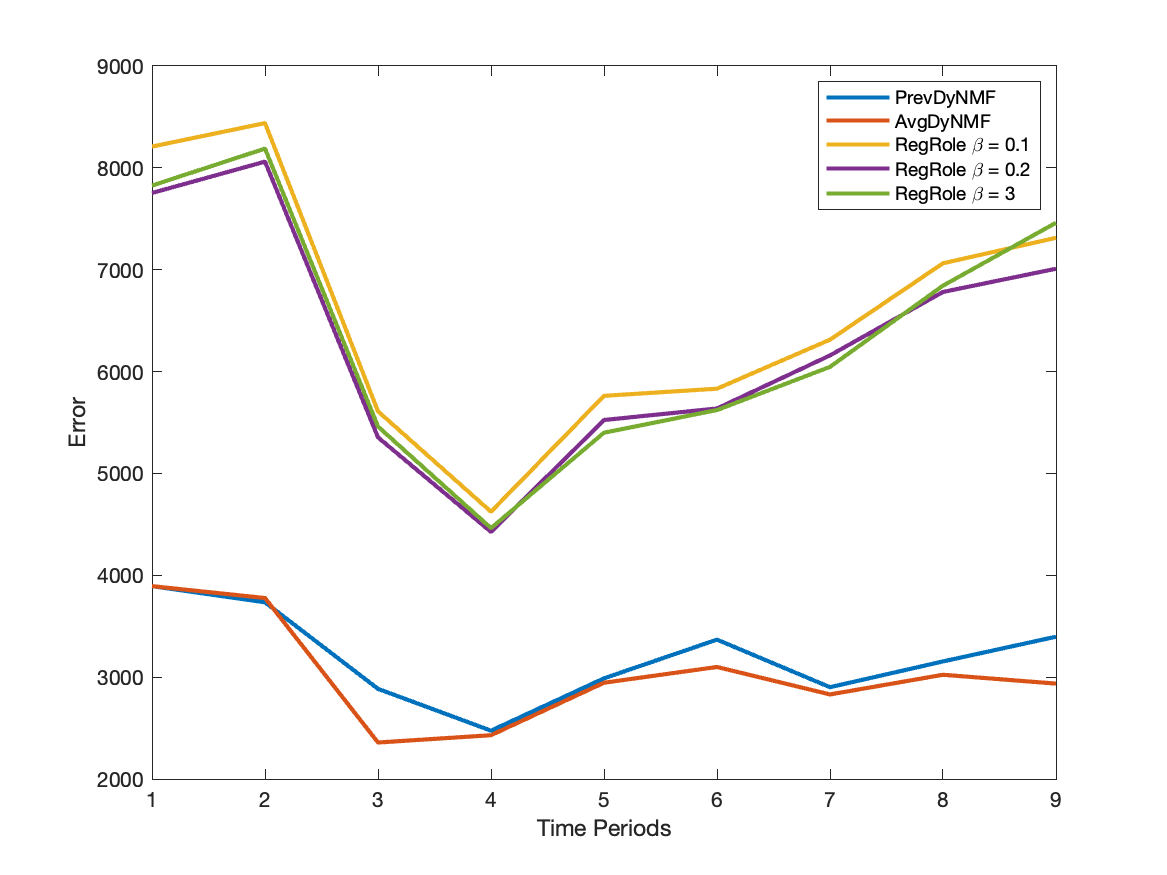}
\caption{Comparison of prediction errors of the proposed method with various different $\beta$ values versus DyNMF (using both the previous and an averaged transition matrix) for the Facebook dataset with the engineered features on the left and the automatic feature generation on the right.  In the left figure the beta values are the middle to bottom curves and that of DyNMF curves appear at the top.  For the automatic feature generation, the opposite pattern is seen.}
  \label{fig:facebookallpredict}
\end{figure}

\begin{table}[]
\centering
\resizebox{0.98\textwidth}{!}{
\begin{tabular}{llcccccc}
\hline
 & & Scratch & Enron & Facebook & Reality & Slashdot & Synthetic \\
\hline
\multirow{6}{*}{\(\text{DyNMF}\ \left.\begin{array}{c}
    \vphantom{A}\\
    \vphantom{A}\\
    \vphantom{A}\\
    \vphantom{A}\\
    \vphantom{A}\\
    \vphantom{A}
    \end{array}\right.\)}
    &Prev Max & 8.6 & 3.4 & 105.2 & 4.9 & 12036.7 & 35.6 \\
&Prev Min & 5.2 & 0.5 & 25.3 & 4.0 & 171.8 & 14.1 \\
&Prev Mean & 7.3 & 2.1 & 61.6 & 4.3 & 2559.7 & 22.7 \\
\cline{2-8}
&Avg Max & 8.8 & 3.0 & 99.2 & 4.7 & 6014.9 & 33.9 \\
&Avg Min & 5.1 & 0.6 & 25.6 & 3.8 & 178.6 & 12.4 \\
&Avg Mean & 7.2 & 1.9 & 60.0 & 4.1 & 3350.1 & 19.3 \\
\hline
\multirow{9}{*}{\(\text{RegRole}\ \left.\begin{array}{c}
    \vphantom{A}\\
    \vphantom{A}\\
    \vphantom{A}\\
    \vphantom{A}\\
    \vphantom{A}\\
    \vphantom{A}\\
    \vphantom{A}\\
    \vphantom{A}\\
    \vphantom{A}
    \end{array}\right.\)}
    &$\beta = 0.1$ Max & 5.1 & 0.6 & 59.5 & 2.7 & 189.0 & 6.5 \\
&$\beta = 0.1$ Min & 4.0 & 0.1 & 19.0 & 2.0 & 158.1 & 3.5 \\
&$\beta = 0.1$ Mean & 4.4 & 0.4 & 38.1 & 2.4 & 173.7 & 5.1 \\
\cline{2-8}
&$\beta = 0.8$ Max & 6.0 & 0.2 & 8.7 & 1.9 & 21.7 & 5.5 \\
&$\beta = 0.8$ Min & 4.7 & 0.0 & 2.3 & 1.5 & 16.6 & 2.1 \\
&$\beta = 0.8$ Mean & 5.2 & 0.1 & 5.0 & 1.7 & 19.2 & 3.4 \\
\cline{2-8}
&$\beta = 3.0$ Max & 2.1 & 0.1 & 4.9 & 1.6 & 25.8 & 5.8 \\
&$\beta = 3.0$ Min & 1.4 & 0.0 & 1.3 & 1.3 & 19.6 & 2.6 \\
&$\beta = 3.0$ Mean & 1.7 & 0.1 & 2.7 & 1.5 & 22.6 & 3.5 \\
\hline
\end{tabular}}   \caption{Summary statistics for prediction error on {six different datasets} for RegRole and DyNMF using engineered features.  Notice that RegRole has lower prediction errors across all of the datasets.}
    \label{tab:rta}
\end{table}

\begin{table}[]
\centering
\begin{tabular}{clccccc}
\hline
& & Facebook & Reality & Enron & Synthetic & Slashdot \\
\hline
\multirow{6}{*}{\(\text{DyNMF}\ \left.\begin{array}{c}
    \vphantom{A}\\
    \vphantom{A}\\
    \vphantom{A}\\
    \vphantom{A}\\
    \vphantom{A}\\
    \vphantom{A}
    \end{array}\right.\)}&Prev Max  & 3893.4 & 12358.3 & 608.8 & 3463.5 & 23368.9 \\
&Prev Min  & 2474.6  & 2520.8 & 4.2   & 1281.2 & 19139.9 \\
&Prev Mean & 3199.7 & 7165.0 & 146.6 & 1784.1 & 21584.1 \\
\cline{2-7}
&Avg Max   & 3893.4 & 12358.3 & 396.7  & 3463.5 & 23527.3 \\
&Avg Min   & 2358.2 & 1787.3 & 4.4   & 1034.7 & 13046.9 \\
&Avg Mean  & 3032.6 & 6710.4 & 122.7  & 1686.1 & 18494.5 \\
\hline
\multirow{9}{*}{\(\text{RegRole}\ \left.\begin{array}{c}
    \vphantom{A}\\
    \vphantom{A}\\
    \vphantom{A}\\
    \vphantom{A}\\
    \vphantom{A}\\
    \vphantom{A}\\
    \vphantom{A}\\
    \vphantom{A}\\
    \vphantom{A}
    \end{array}\right.\)}
    &$\beta=0.1$ Max  & 8151.9 & 17813.1 & 301.3 & 1130.2 & 89000.1 \\
&$\beta=0.1$ Min  & 4484.4 & 5696.7   & 2.3   & 321.1  & 68791.5 \\
&$\beta=0.1$ Mean & 6366.3 & 11789.3 & 76.7  & 568.5  & 77542.4 \\
\cline{2-7}
&$\beta=0.2$ Max  & 8439.3 & 27458.6 & 328.7 & 950.0 & 71383.4 \\
&$\beta=0.2$ Min  & 4624.1 & 8227.6  & 2.3   & 277.8 & 54840.6 \\
&$\beta=0.2$ Mean & 6573.9 & 18619.3 & 83.9  & 486.6 & 62146.5 \\
\cline{2-7}
&$\beta=0.3$ Max  & 7524.7 & 18886.3 & 276.3 & 1103.5 & 65394.2 \\
&$\beta=0.3$ Min  & 4148.3 & 6400.2  & 1.8   & 271.8  & 51467.6 \\
&$\beta=0.3$ Mean & 5874.2 & 12251.0 & 66.3  & 493.5  & 57625.9 \\
\hline
\end{tabular}\caption{Summary statistics for prediction error on {six different datasets} for RegRole and DyNMF using automatically generated features.  Notice that RegRole has lower prediction errors for some of the datasets.}
    \label{tab:rta_AUTO}
\end{table}

\subsection{Experiment 3: Comparing transition matrices}
We next compare the transition matrix from the proposed algorithm to the transition matrices identified by DyNMF.  Similar to DyNMF, we make {no additional restrictions} on our transition matrix in the model besides nonnegativity. Since the magnitude of the values in the transition matrix may vary for different time periods for DyNMF as well as for our algorithm, we do normalize the transition matrix to make it stochastic {so the column entries sum to 1 before conduct comparison}.  We demonstrate the difference in the trace of the transition matrix {using} the Facebook data set {as an example} in Fig~\ref{fig:facebookTrans}.  In Tables~\ref{tab:transition} and ~\ref{tab:transition_AUTO} we display summary statistics for all our data sets.  We observe that the trace values are larger for RegRole which indicates individuals are more likely to stay in their role, which is as expected when a penalty is applied for changing roles.

\begin{figure}
    \centering
    \includegraphics[width=0.48\textwidth]{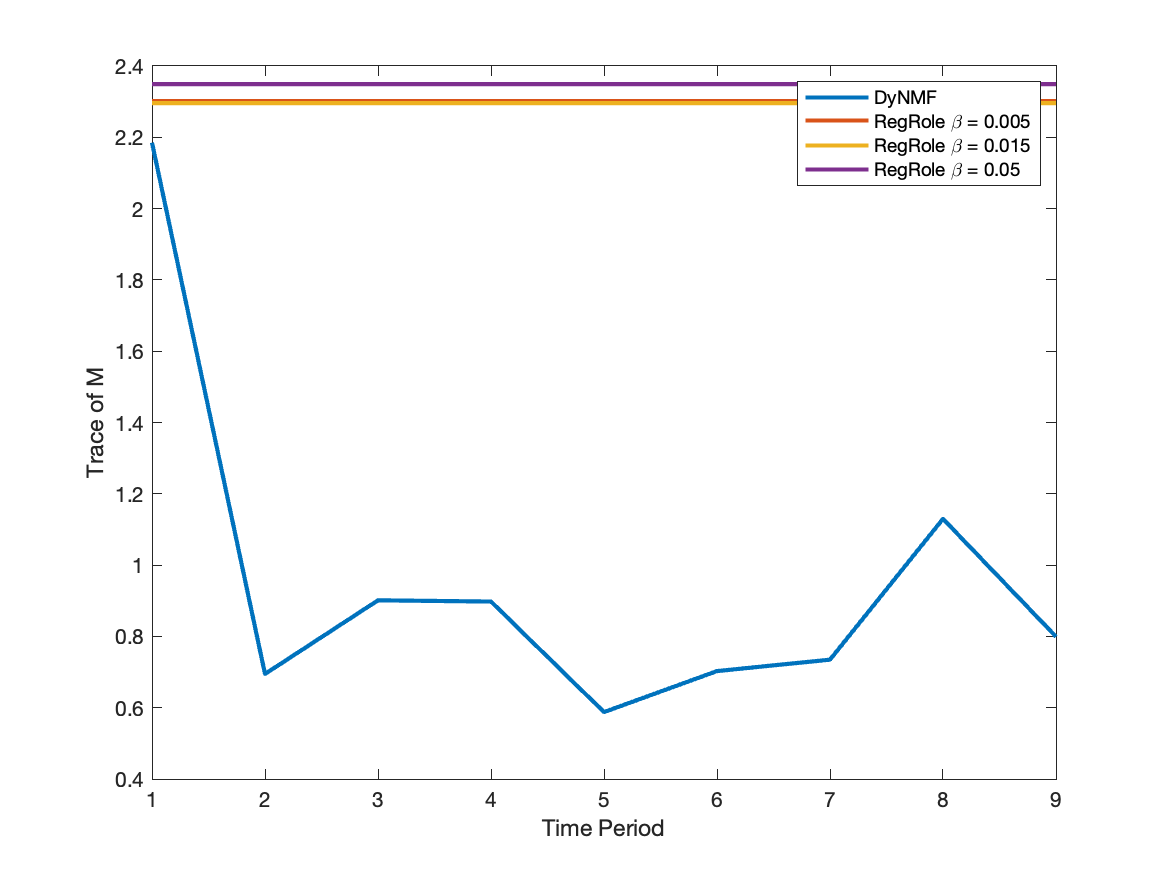}\hfill
    \includegraphics[width=0.48\textwidth]{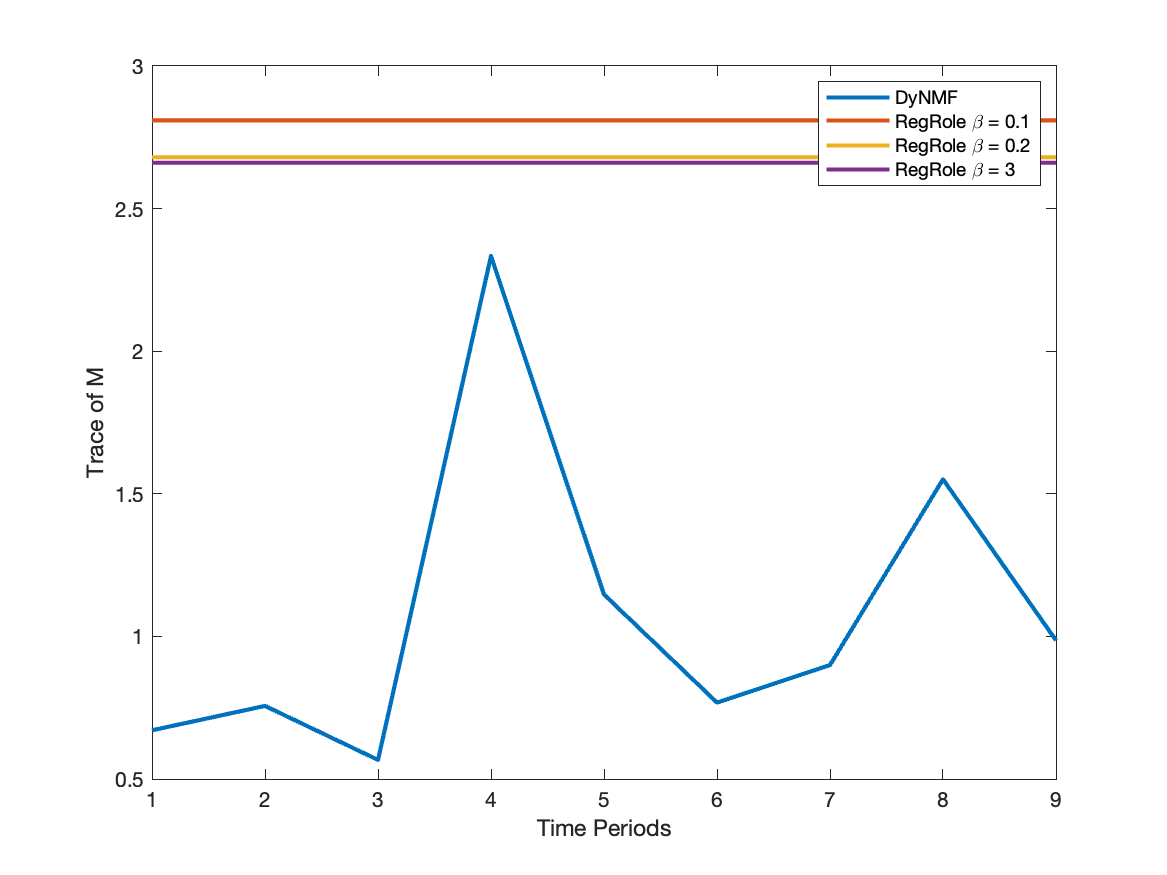}
    \caption{Trace of transition matrix for the Facebook dataset with engineered features on the left and automatically generated features on the right.  Observe that the RegRole matrices all had higher trace values than DyNMF. This  indicates our proposed scheme leads to a more stable system (i.e., less changes at each time snapshot).}
    \label{fig:facebookTrans}
\end{figure}

\begin{table}[]
    \centering
    \resizebox{0.98\textwidth}{!}{
\begin{tabular}{llcccccc}
\hline
& & Scratch & Enron & Facebook & Reality & Slashdot & Synthetic \\
\hline
\multirow{3}{*}{\(\text{DyNMF}\ \left.\begin{array}{c}
    \vphantom{A}\\
    \vphantom{A}\\
    \vphantom{A}
    \end{array}\right.\)}&Mean& 1.0 & NaN & 0.9 & 1.0 & 0.9 & 1.2 \\
&Min  & 0.6 & 0.1 & 0.5 & 0.3 & 0.2 & 0.5 \\
&Max& 1.4 & 2.4 & 1.6 & 2.0 & 1.7 & 2.5 \\
\hline
\multirow{3}{*}{\(\text{RegRole}\ \left.\begin{array}{c}
    \vphantom{A}\\
    \vphantom{A}\\
    \vphantom{A}
    \end{array}\right.\)}&$\beta=0.1$ & 1.1 & 2.7 & 2.3 & 3.3 & 2.3 & 2.2 \\
&$\beta=0.8$ & 1.1 & 2.6 & 2.0 & 3.4 & 2.3 & 1.8 \\
&$\beta=3.0$ & 1.1 & 2.7 & 2.3 & 3.5 & 2.6 & 2.2 \\
\hline
\end{tabular}}   \caption{Summary Statistics for trace calculations with engineered data of {transition matrices between DyNMF (top panel) and the proposed one}. Observe that in all cases the mean DyNMF trace value is lower than the trace value for our algorithm.  This indicates a more stable scenario, i.e., less transitions between roles. }
    \label{tab:transition}
\end{table}

\begin{table}[]
    \centering
\begin{tabular}{llccccc}
\hline
 & & Facebook & Reality & Enron & Synthetic & Slashdot \\
\hline
\multirow{3}{*}{\(\text{DyNMF}\ \left.\begin{array}{c}
    \vphantom{A}\\
    \vphantom{A}\\
    \vphantom{A}
    \end{array}\right.\)}&Mean& 1.1 & 0.8 & 1.0 & NaN    & 1.2 \\
&Min & 0.6 & 0.4 & 0.0 & 0.5 & 0.8 \\
&Max & 2.3 & 1.9 & 3.3 & 1.7 & 1.6 \\
\hline
\multirow{3}{*}{\(\text{RegRole}\ \left.\begin{array}{c}
    \vphantom{A}\\
    \vphantom{A}\\
    \vphantom{A}
    \end{array}\right.\)}&$\beta=0.1$  & 2.8 & 2.1 & 2.9 & 1.8 & 1.2 \\
&$\beta=0.2$ & 2.7 & 2.7 & 2.9 & 1.7 & 0.8 \\
&$\beta=0.3$  & 2.7  & 2.4 & 3.1 & 1.6 & 0.9  \\
\hline
\end{tabular}
\caption{Summary Statistics for trace calculations with automatically generated features.  We note that in all cases the trace values corresponding to the sampled $\beta$ values, are greater than the mean values for DyNMF, indicating a more stable scenario.}
    \label{tab:transition_AUTO}
\end{table}

 \subsection{Experiment 4: Role prediction with a known ground truth}
 For our next set of experiments we generate data to test the predictive power of our method when a ground truth is known.  To generate a synthetic data with ground truth we first run our proposed model on the given data (the number of time snapshots varies between datasets and is given in Table~\ref{tab:datasetinfo}) to determine a universal role matrix $V$, a universal transition matrix $M$ and a starting distribution matrix $U$ (i.e., $U_0$ ) that describes the distribution of roles between nodes for our initial time conditions. Letting $\hat{U}_0=U$ be our ground truth starting matrix, we define ground truth role distribution matrices for each time step as $\hat{U}_i = \hat{U}_0M^i$ for $i = 1, \ldots, 20,$ and we set $X_i = \hat{U}_iV+\varepsilon$. {$X_i$ is then treated as synthetic data with ground truth.} We then {take $X_i$ as input and} run RegRole with transition predictions algorithm to find an estimated $U_t$ denoted as $\hat{U}_t$  for each of the time snap shots.  Since each row of $U$ represents the role distribution for an individual we measure the distribution estimation error by determining the Frobenius norm $||U_t - \hat{U}_t||_F$. A visualization of the results are shown for the Facebook dataset in Fig~\ref{fig:FroFacebookSyn}.  As can be seen RegRole performs better (for the displayed values of $\beta$ and many other values of $\beta$) than DyNMF.  We note that the reasonable unnormalized values of $\beta$ are higher for the engineered features than the automatic features.  This is to be expected since we normalize $\beta$ by multiply by $r=6$ and dividing by the number of features.  Since the number of automatically generated features is more than 10 time higher than the number of engineered features we expect the valid good $\beta$ values to be proportionately larger.  We also determine the KL divergence between $U_t$ and $\hat{U}_t$ which is visualized in Fig~\ref{fig:real_pred_kl}.  We note that the KL divergence is not necessarily lower for all of the chosen $\beta$ values, but most are at least equivalent to the DyNMF algorithm.  The data for all of the datasets are shown in Tables~\ref{tab:SyntheticDataExperimentsEng} and~\ref{tab:SyntheticDataExperimentsAuto}.

 \begin{figure}[htbp]
  \centering
 \includegraphics[width=0.48\textwidth]{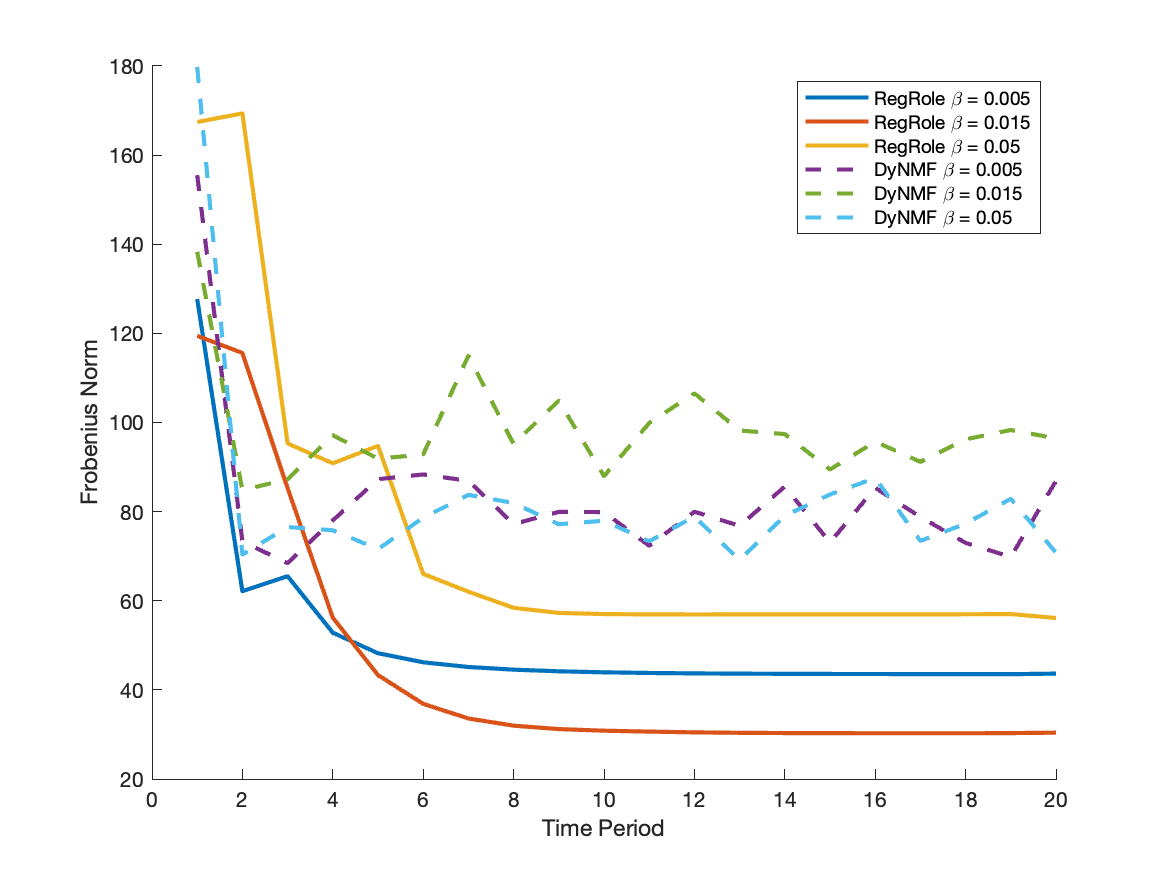}\hfill
 \includegraphics[width=0.48\textwidth]{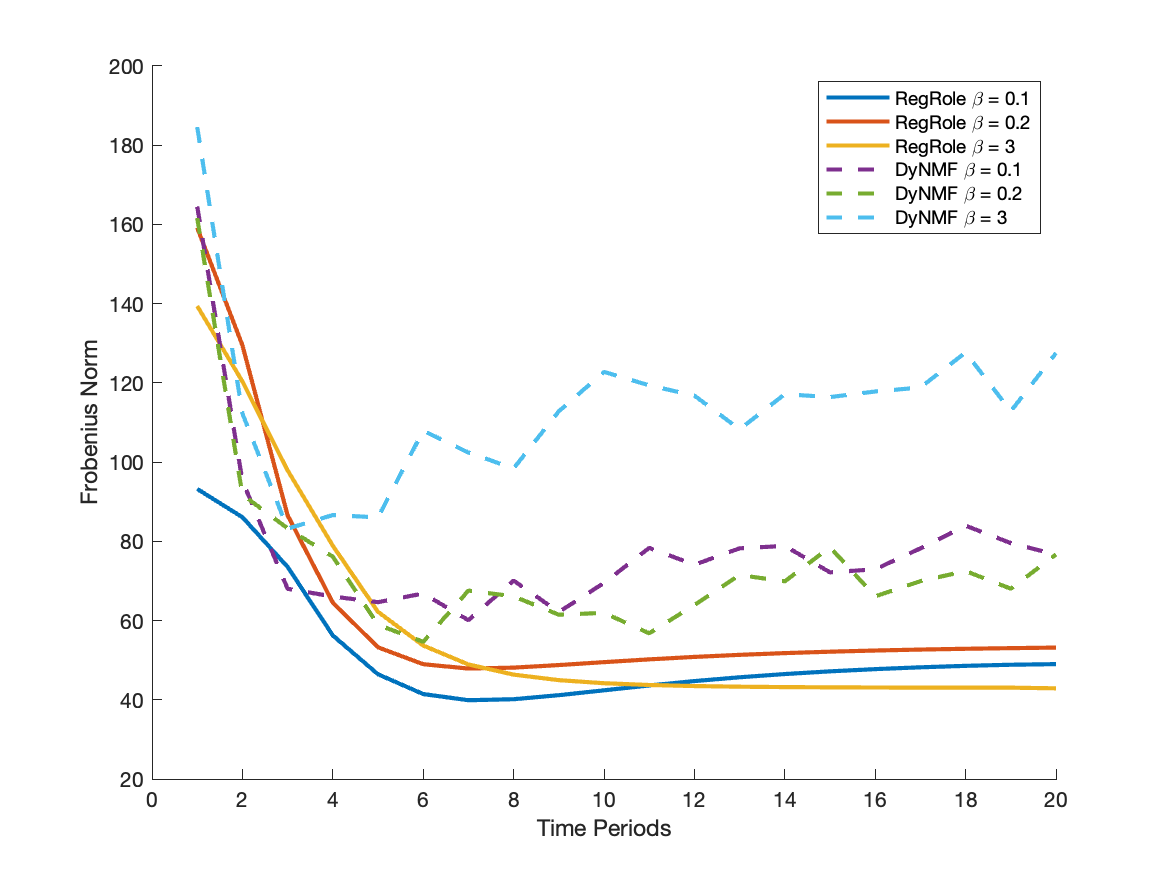}
  \caption{{Error in estimation of $U_t$} using synthetically generated data from the Facebook dataset as measured by the Frobenius norm: left panel: engineered features; right panel: automatically generated features. Observe that the error decreases initially and stabilizes for RegRole.  DyNMF also experiences an initial rapid decrease, but does not stabilize.  In all cases the error for the RegRole in lower than for DyNMF. The $\beta$ values indicate the $\beta$ values used when computing the ground truth data and for the RegRole results the value of $\beta$ used when computing $U_t$.  DyNMF does not use $\beta$ when determining $U_t$, the $\beta$ values are shown to clarify which synthetic dataset was utilized.}
  \label{fig:FroFacebookSyn}
\end{figure}

\begin{figure}[htbp]
  \centering
 \includegraphics[width=0.48\textwidth]{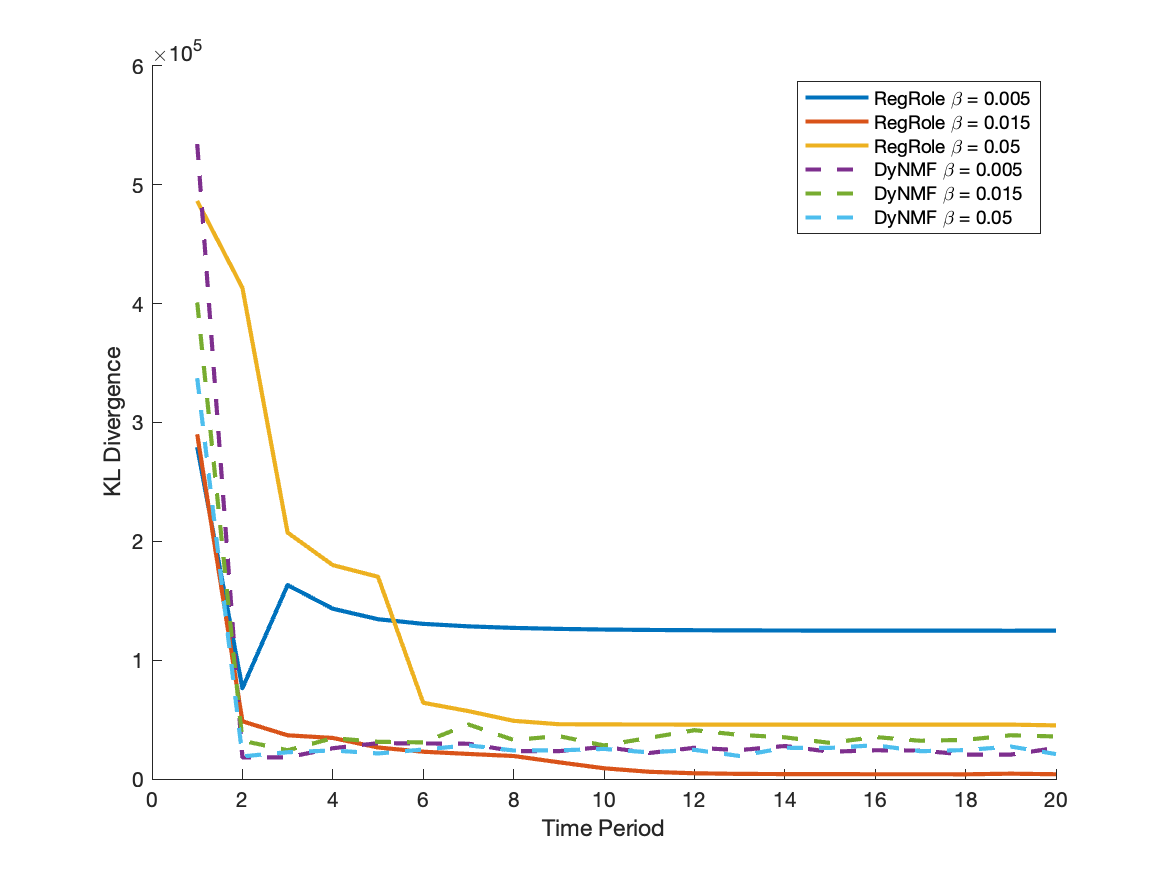}\hfill
 \includegraphics[width=0.48\textwidth]{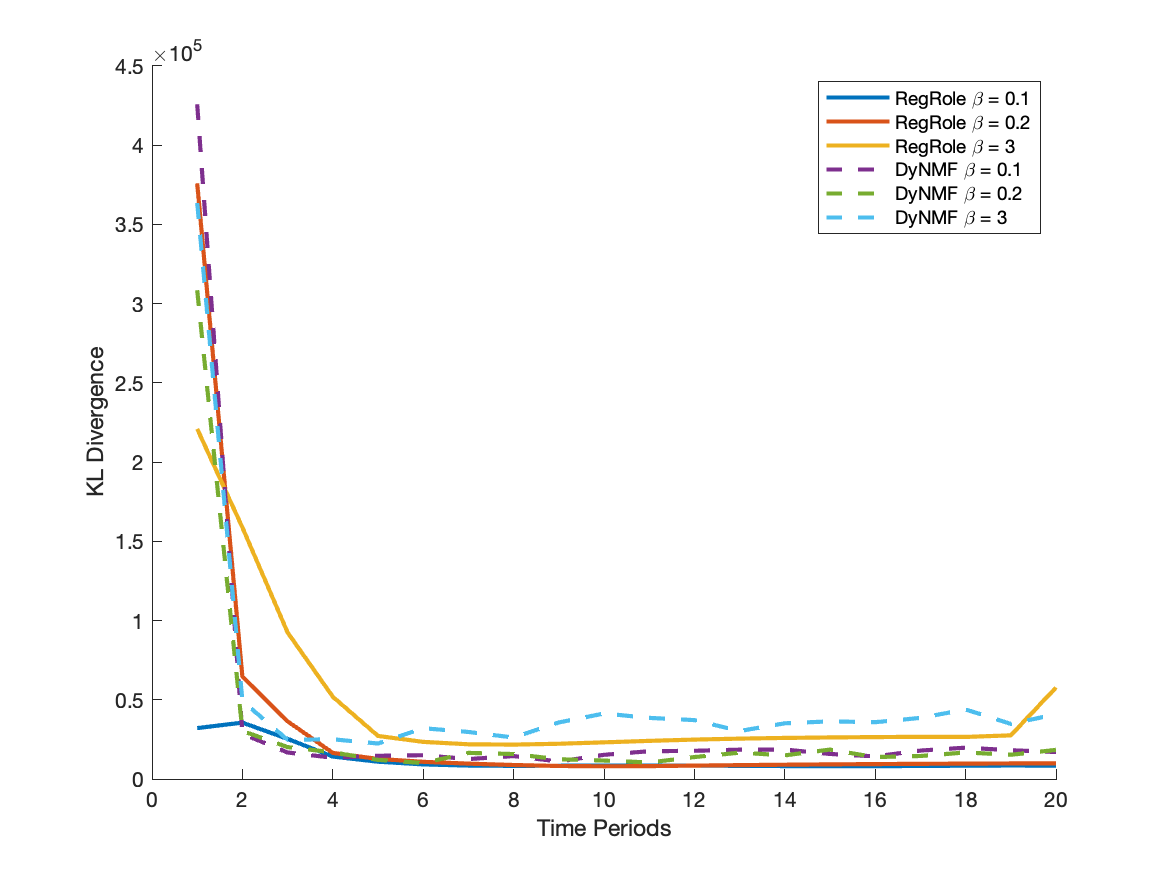}
  \caption{{Error in estimation of $U_t$ using synthetically generated data from the Facebook dataset as measured with KL Divergence. For the automatically generated features (right panel) the KL divergence is almost equivalent between the two algorithms, with two of the selected $\beta$ values giving better results than DyNMF.  For the engineered data (left panel), the orange one (RegRole with $\beta = 0.015$) is the best. This underlines the importance of selecting the appropriate $\beta$ values.}} 
\end{figure}

  \begin{table}[]
    \centering
    \centering
\begin{tabular}{lcccccc}
\hline
 & Scratch & Enron & Facebook & Reality & Slashdot & Synthetic \\
\hline
\multicolumn{7}{c}{$\beta=0.1$} \\
Max RegRole & 64.0 & 7.5 & 157.8 & 33.5 & 223.7 & 1.9 \\
Min RegRole & 11.4 & 1.3 & 56.1 & 19.0 & 93.0 & 0.7 \\
Mean RegRole & 14.5 & 1.7 & 84.5 & 21.8 & 118.0 & 1.0 \\
Max DyNMF & 62.6 & 7.4 & 167.1 & 49.2 & 176.6 & 5.8 \\
Min DyNMF & 47.2 & 1.8 & 72.1 & 23.1 & 74.8 & 2.3 \\
Mean DyNMF & 54.9 & 2.6 & 85.1 & 27.1 & 95.3 & 3.2 \\
\hline
\multicolumn{7}{c}{$\beta=0.8$} \\
Max RegRole & 51.8 & 5.4 & 158.2 & 56.2 & 193.4 & 6.0 \\
Min RegRole & 10.0 & 0.9 & 111.7 & 32.7 & 115.4 & 5.4 \\
Mean RegRole & 13.0 & 1.2 & 144.7 & 35.2 & 145.4 & 5.7 \\
Max DyNMF & 77.5 & 6.0 & 113.4 & 42.5 & 154.4 & 6.0 \\
Min DyNMF & 49.4 & 1.8 & 45.1 & 19.0 & 96.2 & 2.4 \\
Mean DyNMF & 57.9 & 2.3 & 60.5 & 26.7 & 132.6 & 3.7 \\
\hline
\multicolumn{7}{c}{$\beta=3.0$} \\
Max RegRole & 59.7 & 7.8 & 223.2 & 48.5 & 185.2 & 8.5 \\
Min RegRole & 29.7 & 0.6 & 141.7 & 13.6 & 78.9 & 2.0 \\
Mean RegRole & 44.8 & 1.2 & 156.7 & 32.6 & 152.5 & 3.0 \\
Max DyNMF & 68.9 & 7.8 & 105.2 & 48.5 & 210.4 & 5.7 \\
Min DyNMF & 28.6 & 1.8 & 69.7 & 18.1 & 73.4 & 3.0 \\
Mean DyNMF & 37.9 & 2.7 & 96.2 & 23.2 & 100.4 & 4.1 \\
\hline
\end{tabular}
    \caption{Summary data for experiments run with engineered features and synthetic data, i.e., ground truth roles known. Observe that for most data and for both values of $\beta$ RegRole yields lower value for minium, maximum, and mean values.}
    \label{tab:SyntheticDataExperimentsEng}
\end{table}
   \begin{table}[]
    \centering
    \centering
    \begin{tabular}{lccccc}
\hline
 & Facebook & Reality & Enron & Synthetic & Slashdot \\
\hline
\multicolumn{6}{c}{$\beta=0.1$} \\
Max RegRole  & 93.2 & 85.4 & 8.2 & 7.0 & 99.7 \\
Min RegRole  & 39.9 & 39.4 & 3.6 & 1.5 & 74.2 \\
Mean RegRole & 51.5 & 49.3  & 5.7 & 2.2 & 89.6 \\
Max DyNMF    & 164.6 & 85.0 & 8.2 & 7.2  & 211.3 \\
Min DyNMF    & 60.1  & 21.8 & 2.8 & 3.0  & 71.3 \\
Mean DyNMF   & 78.0  & 31.4 & 3.6 & 4.2 & 88.6 \\
\hline
\multicolumn{6}{c}{$\beta=0.2$} \\
Max RegRole  & 105.5 & 55.8 & 8.5 & 1.7 & 228.0 \\
Min RegRole  & 54.9  & 22.2 & 2.2  & 1.2 & 98.6 \\
Mean RegRole & 67.8   & 32.3 & 3.2 & 1.5 & 116.5 \\
Max DyNMF    & 142.7 & 71.7 & 9.4 & 7.1 & 222.0 \\
Min DyNMF    & 56.3  & 24.8 & 3.1 & 3.2 & 53.3 \\
Mean DyNMF   & 76.2  & 33.1 & 4.0 & 4.6 & 73.3 \\
\hline
\multicolumn{6}{c}{$\beta=0.3$} \\
Max RegRole  & 165.0 & 50.1 & 9.6 & 9.4 & 206.1 \\
Min RegRole  & 73.8  & 14.7 & 4.1 & 1.9 & 56.8 \\
Mean RegRole & 95.0  & 22.5 & 7.2 & 2.4 & 69.0 \\
Max DyNMF    & 156.9 & 69.8 & 9.3 & 8.7  & 219.1 \\
Min DyNMF    & 55.6  & 25.1 & 2.5 & 2.8 & 55.7 \\
Mean DyNMF   & 86.1  & 32.4 & 3.4 & 4.2 & 75.2 \\
\hline
\end{tabular}
   \caption{Summary data for experiments run with automatic feature generation and synthetic data, i.e., ground truth roles known. Observe that for most data and for both values of $\beta$ RegRole yields lower value for minium, maximum, and mean values.}
    \label{tab:SyntheticDataExperimentsAuto}
\end{table}

 \subsection{Experiment 5: RegRole for forward predictions}
 Finally we consider the predictive power of RegRole to make accurate forward predictions of an individual's role make up.  We perform two tests.  For the first test, we utilize a similar technique to the prior test, we train our model on real data (with a fixed beta) and generate twenty time steps of ground truth data.  We then use the first 15 time snapshots to calculate unknowns and the last five to compare the relative Frobenius norm (see Fig. \ref{fig:Facebook_syn_pred_frob}) and KL divergence (see Fig. \ref{fig:Facebook_syn_pred_KL}) between the predicted results and the ground truth. {One can see that the difference does not change much as time period increases.}

The second way we test the predictive power of our model is to use the actual datasets. Given our data set $X$ we divide our data set into a training period $X_R$ and a test period $X_S.$  Using the training data we calculate the universal role matrix $V$, the universal transition matrix $M$, and a role composition matrix $U_t$ for each time in the training set. {The prediction of $U_t$ on time snapshots in the test period is obtained by keeping multiplying by the transition matrix $M$. Due to the lack of ground truth $U_t$ to compare with, we use the following approach to calculate $U_t$.}  For the time step in the test period we calculate a nonnegative factorization $X_t = U_tV_t$ for each time $t$.  We then reorder the vectors in $V_t$ so they match the role vectors in the universal role matrix $V$.  We note that the vectors in $V_t$ and $V$ are not identical and choose the ordering that minimizes $||V-V_t||_F$.  Upon establishing the new ordering of the vectors in $V_t$ we reorder the columns of $U_t$ correspondingly to establish the ground truth.  We 
%summary statistics in Tables~\ref{tab:realdatapred}-\ref{tab:realdatapred_AUTO}, and 
display the results for several values of $\beta$ for the Facebook data in Figs~\ref{fig:real_pred_fro}-\ref{fig:real_pred_kl}. We observe that the prediction error undergoes a significant drop between the initial prediction and subsequent predictions.  This is true across all datasets, not just the Facebook dataset.  We hypothesize that the higher initial error indicates a burn-in period, before our predictions become more accurate.  As time passes the predictive power of our technique begins to fade which is also expected.

\begin{figure}[htbp]
  \centering
 \includegraphics[width=0.48\textwidth]{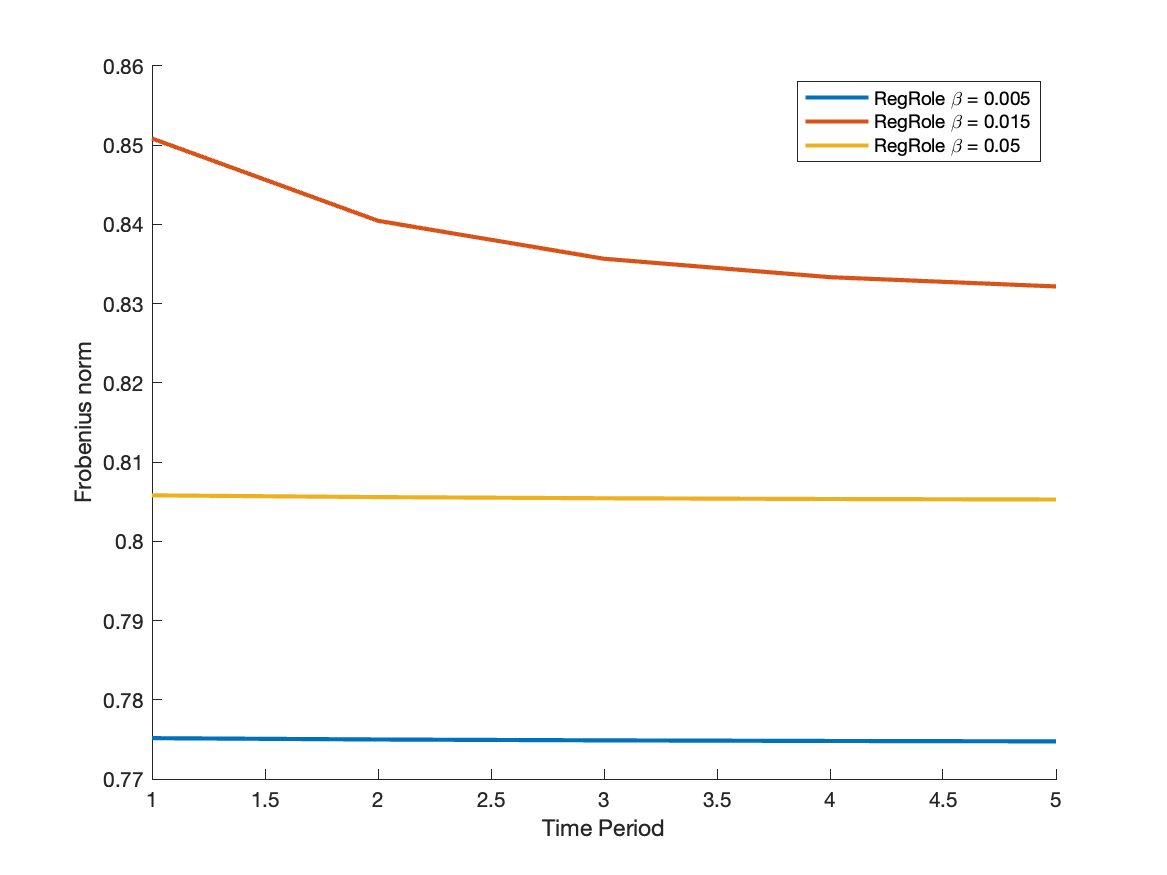}\hfill
 \includegraphics[width=0.48\textwidth]{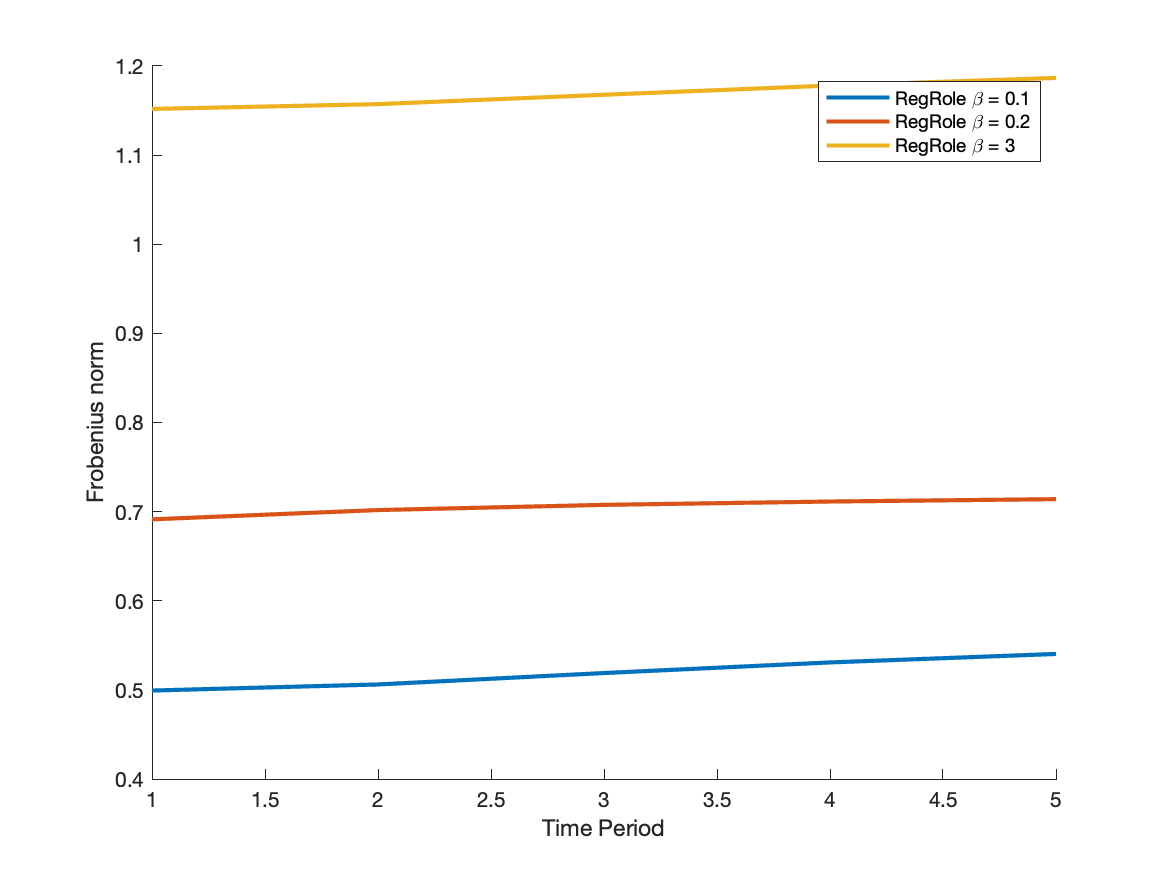}
  \caption{Prediction error using synthetically generated data from the Facebook dataset: left panel: engineered features; right panel: automatically generated features. 20 time snapshots were simulated. {The first 15 was used to calculate unknowns and the last 5 was used for prediction to compare KL divergence of the predicted results and the ground truth.} }
  \label{fig:Facebook_syn_pred_frob}
\end{figure}

\begin{figure}[htbp]
  \centering
 \includegraphics[width=0.48\textwidth]{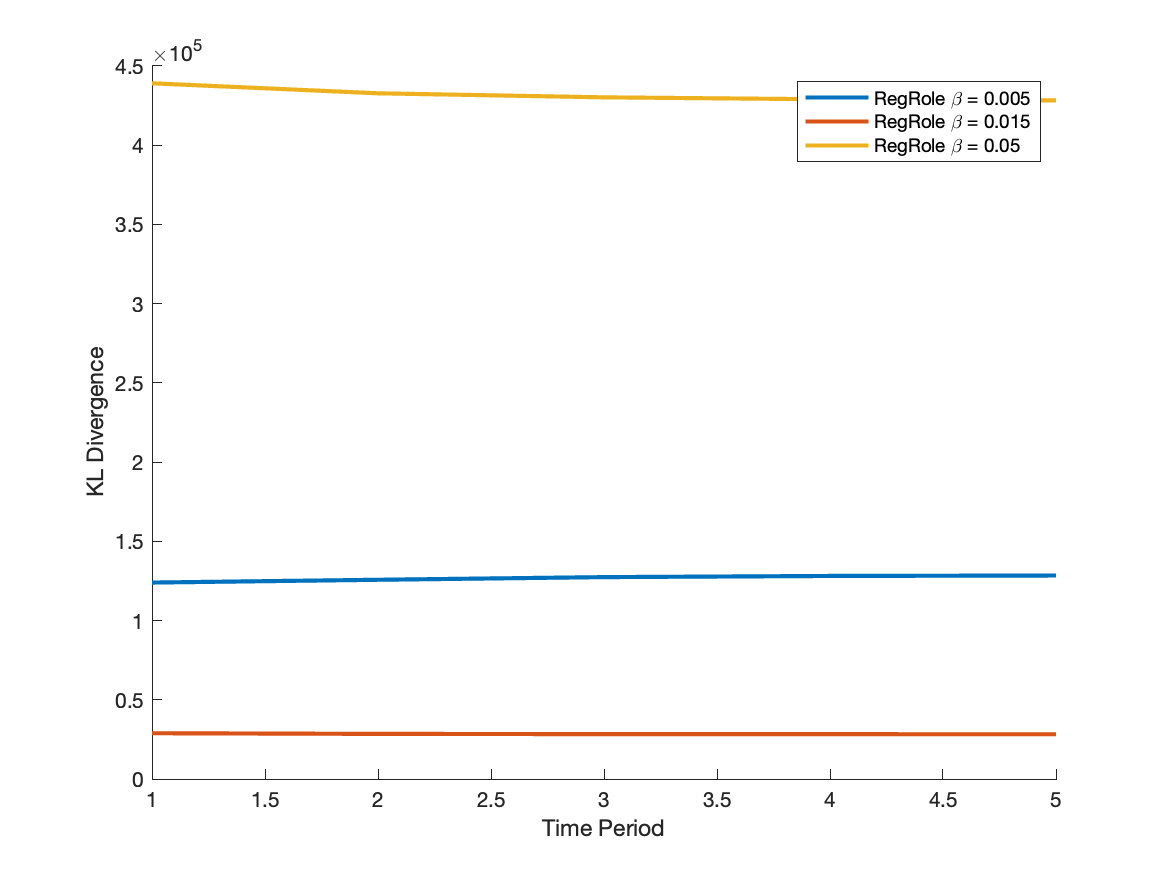}\hfill
 \includegraphics[width=0.48\textwidth]{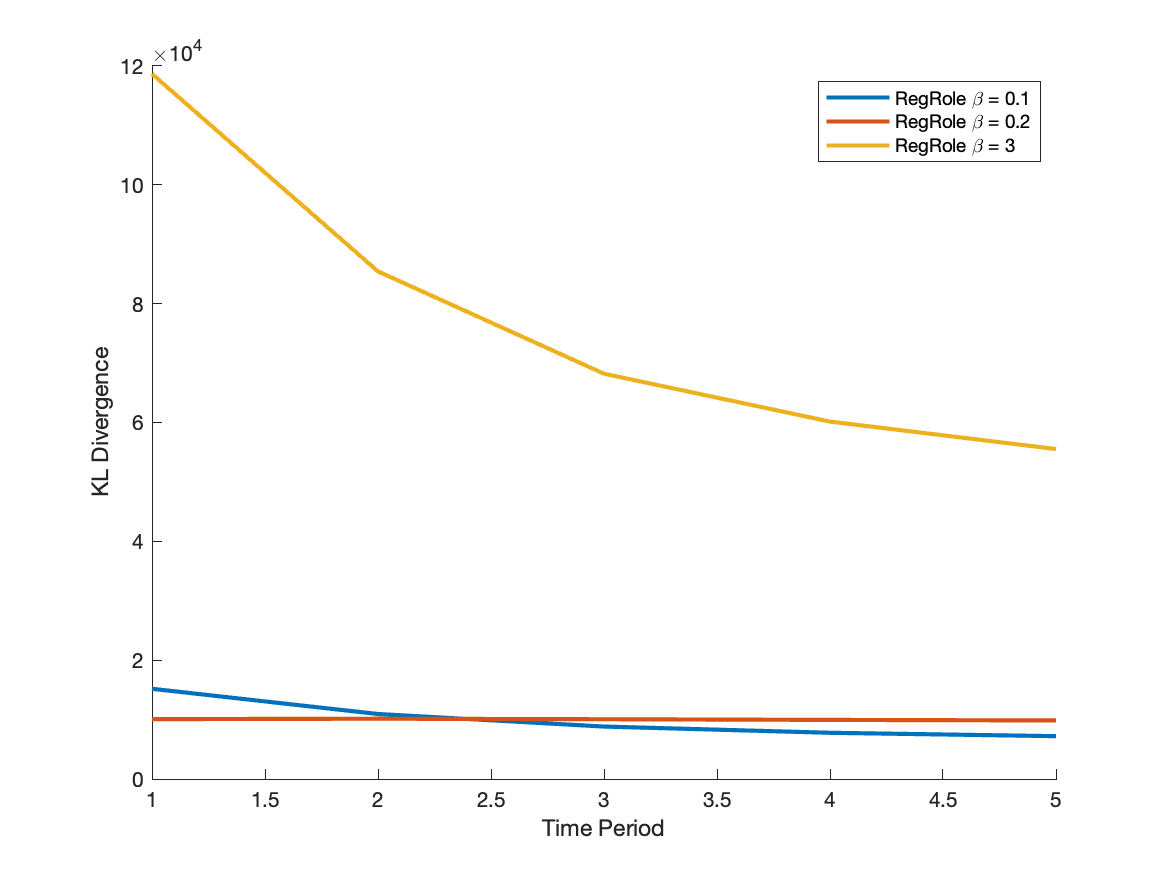}
  \caption{Prediction error using synthetically generated data from the Facebook dataset: left panel: engineered features; right panel: automatically generated features. 20 time snapshots were simulated. The first 15 was used to calculate unknowns and the last 5 was used for prediction to compare KL divergence of the predicted results and the ground truth. }
    \label{fig:Facebook_syn_pred_KL}
\end{figure}

\begin{figure}
    \centering
    \includegraphics[width=0.48\textwidth]{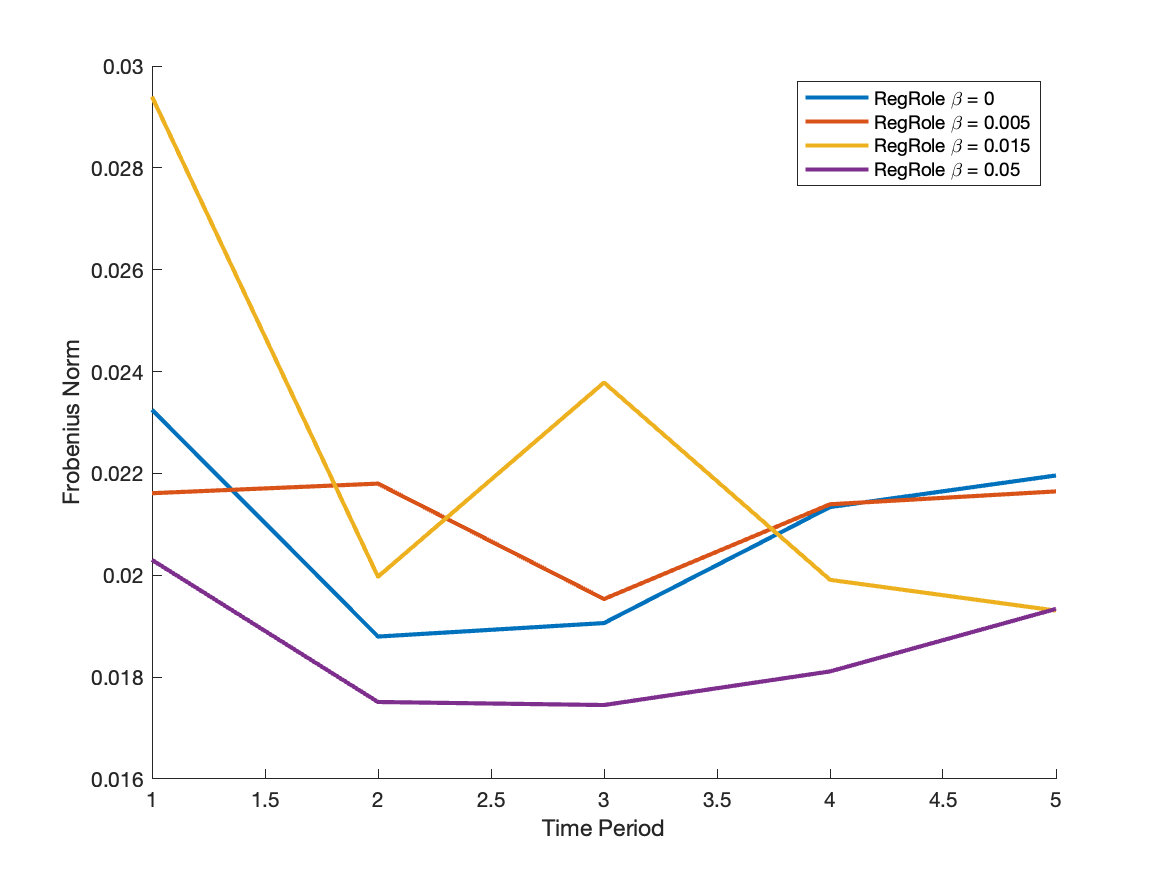}\hfill
 \includegraphics[width=0.48\textwidth]{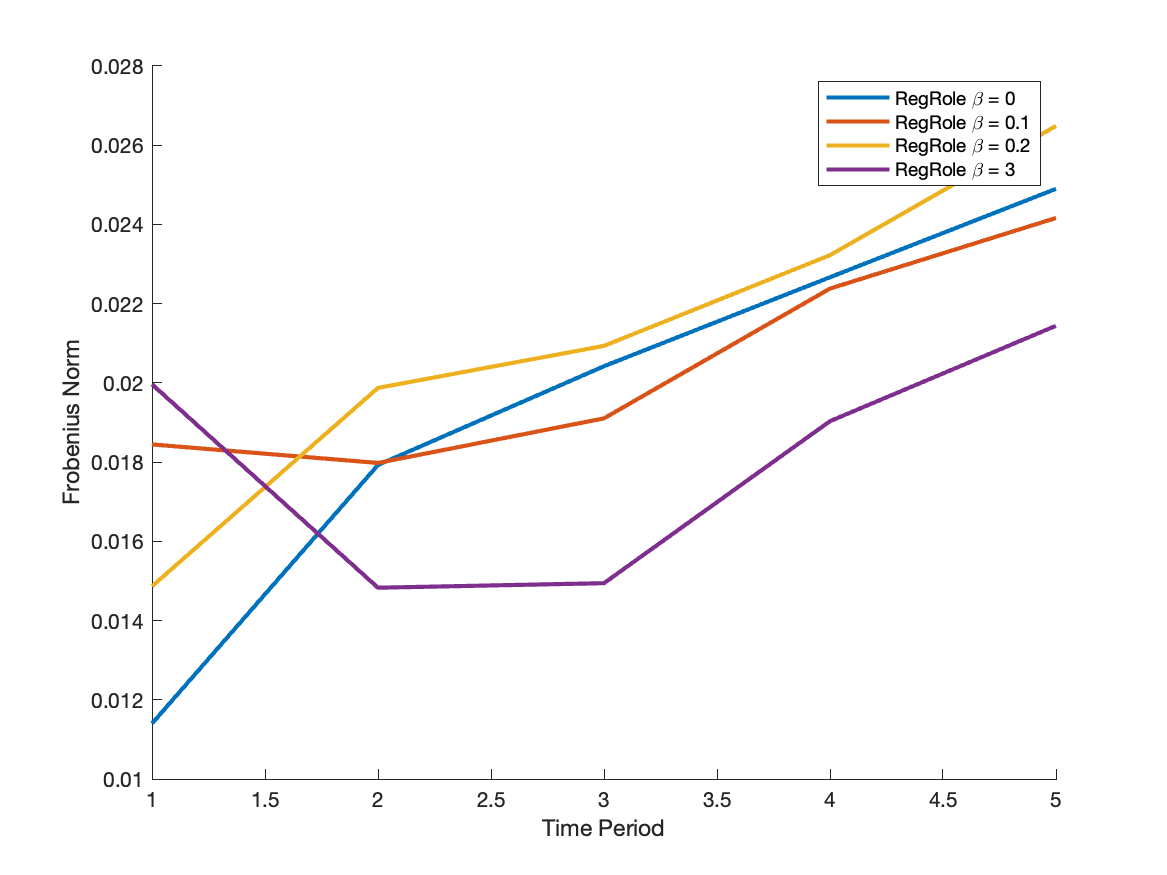}
    \caption{Prediction error as measured by the Frobenius norm using the real Facebook dataset: left panel: engineered features; right panel: automatically generated features.  Our model was trained on the first 5 time periods and predicted on the {next} 5 time periods. Observe that some regularity was better than no regularity (i.e., $\beta = 0$ was generally greater than $\beta \neq 0$).}
    \label{fig:real_pred_fro}
\end{figure}
\begin{figure}
    \centering
    \includegraphics[width=0.5\textwidth]{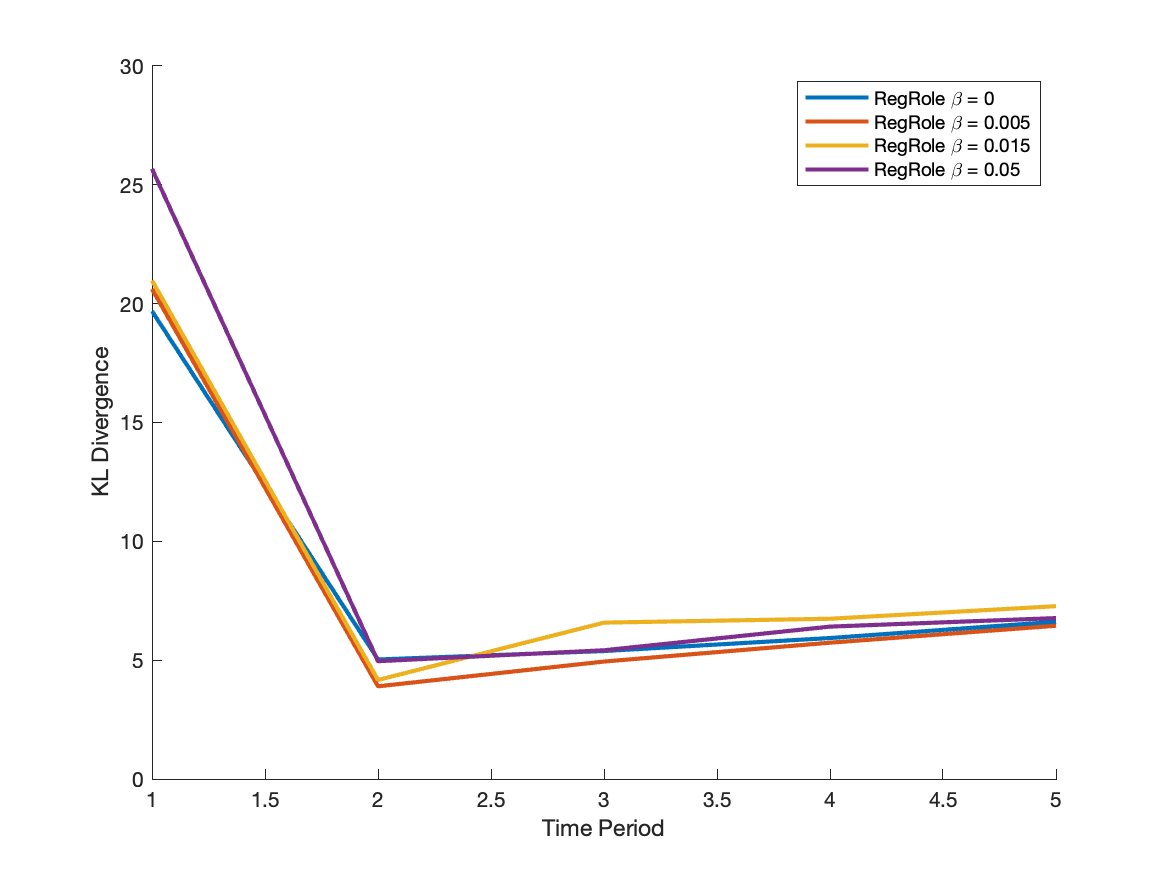}\hfill
 \includegraphics[width=0.48\textwidth]{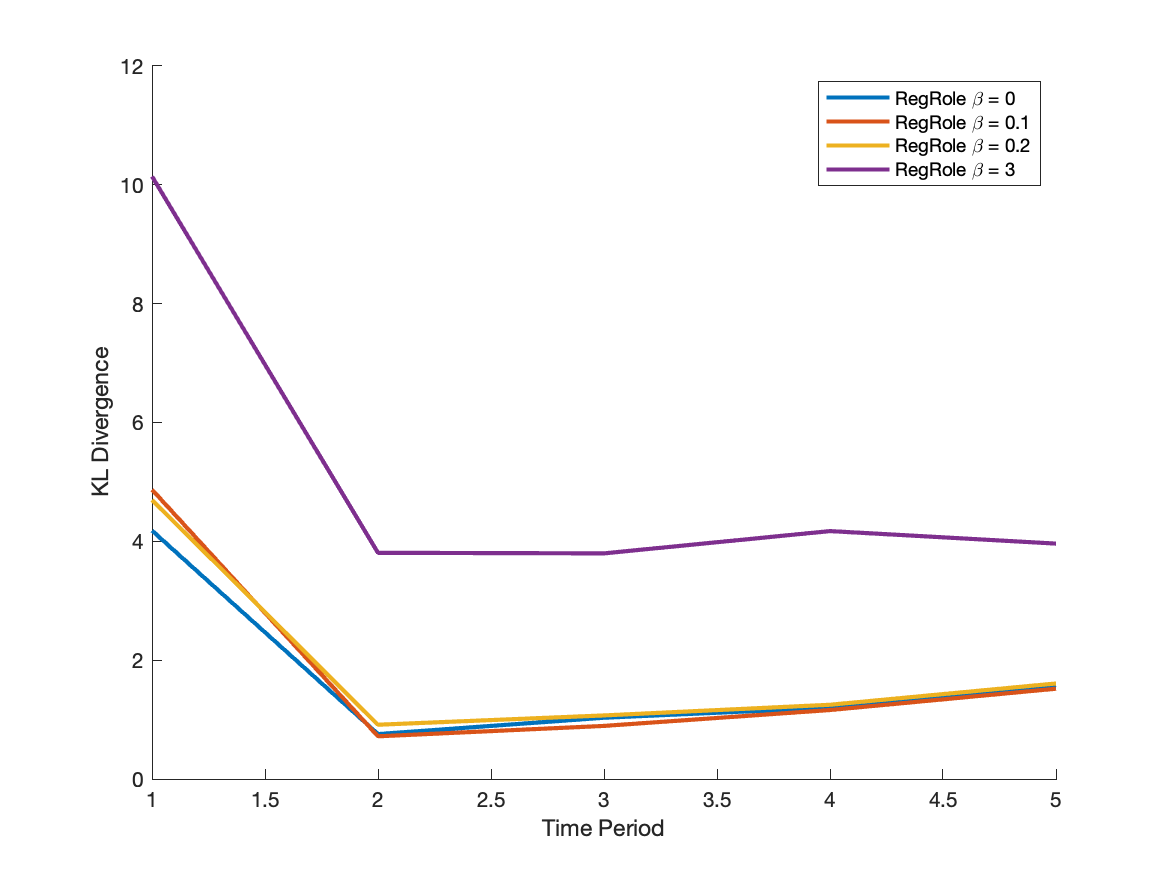}
     \caption{Prediction error as measured by the KL Divergence using the real Facebook dataset: left panel: engineered features; right panel: automatically generated features.  Our model was trained on the first 5 time periods and predicted on the 5 time periods. Observe that some regularity was better than no regularity (i.e., $\beta = 0$ was generally greater than $\beta \neq 0$), although this was not as pronounced as for the Frobenius norm.  }
    \label{fig:real_pred_kl}
\end{figure}

%*****OLD*****

%Then for each time period $t$ in our test set $X_S$ we calculate a ``ground truth'' role distribution by letting $U_t = X_t V^{\dagger}$ where $V^\dagger$ is the Moore-Penrose pseudo inverse.  For each time period $t$ in the test set we let the predicted role distribution $U_t^p = U_{\ell}M^t$ where $U_{\ell}$ is the final role distribution in the training set.
\section{Conclusions and Future Work}\label{sec:con}

In this work, we presented RegRole, a dynamic topic discovery framework for role analysis built on temporally regularized nonnegative matrix factorization. By jointly modeling structural and behavioral features, the proposed approach learns time-aligned role topics while enforcing smooth evolution in node-level role mixtures. This design enables the model to capture meaningful temporal dynamics without sacrificing interpretability or scalability, making it well-suited for large, sparse graphs.

Our empirical results demonstrate clear advantages over existing methods. On synthetic datasets with known ground truth, the framework achieves the lowest prediction error, outperforming DyNMF. Across six real-world datasets, it consistently delivers superior performance compared to state-of-the-art approaches. In addition, the model produces more stable role semantics over time, enhances the detection of structural changes, and improves downstream predictive tasks.

Overall, this work highlights the importance of incorporating temporal regularization into role discovery and provides a robust, scalable solution for analyzing evolving networks. Future work may explore extensions to incorporate richer feature modalities, adaptive temporal constraints, and applications to streaming or real-time graph settings.

\section*{Acknowledgments}
This material is based upon work supported by the National Science Foundation under Grant No. DMS-1929284 while the author was in residence at the Institute for Computational and Experimental Research in Mathematics in Providence, RI, during the ``Learning Temporal Representations of Dynamic Networks for Anomaly Detection and Community Discovery" Colloborate@ICERM program.  We appreciate the collaborative insights offered by Emilee Purvine and Ferhat Arslan.
\bibliography{conf,refs}
\end{document}